\documentclass[10pt,journal,compsoc]{IEEEtran}
\usepackage{comment}

\ifCLASSOPTIONcompsoc
  \usepackage[nocompress]{cite}
\else
  \usepackage{cite}
\fi
\ifCLASSINFOpdf
   \usepackage[pdftex]{graphicx}
\else
   \usepackage[dvips]{graphicx}
\fi
\usepackage{amsmath}

\usepackage[pdftex,dvipsnames]{xcolor}  % Coloured text etc.
\usepackage{multirow}

\usepackage{hyperref}
\usepackage{tikz}
\def\checkmark{\tikz\fill[scale=0.4](0,.35) -- (.25,0) -- (1,.7) -- (.25,.15) -- cycle;}

\usepackage{amssymb}                   % added by jie for math symbols
\usepackage{amsmath}
\usepackage{todonotes}

\newcommand{\major}[1]{\textcolor{black}{#1}}
\usepackage{amsmath,amsfonts,bm}

\def\eqref#1{equation~\ref{#1}}
\def\1{\bm{1}}

\def\mW{{\bm{W}}}

\DeclareMathAlphabet{\mathsfit}{\encodingdefault}{\sfdefault}{m}{sl}
\SetMathAlphabet{\mathsfit}{bold}{\encodingdefault}{\sfdefault}{bx}{n}

\begin{document}
\title{“Understanding Robustness Lottery”: A Geometric Visual Comparative  Analysis of Neural Network Pruning Approaches}

%\title{SpotSDC: Visualizing Silent Data Corruption and Corruption Propagation}

\author{Zhimin Li,
        Shusen Liu,
        Xin Yu,
        Kailkhura Bhavya,
        Jie Cao,
        James Daniel Diffenderfer, 
        Peer-Timo~Bremer,~\IEEEmembership{Member,~IEEE,}
         ~Valerio~Pascucci,
         ~\IEEEmembership{Member,~IEEE}%,~\IEEEmembership{Life~Fellow,~IEEE}% <-this % stops a space
\IEEEcompsocitemizethanks{
\IEEEcompsocthanksitem Zhimin~Li and Valerio~Pascucci are with the Scientific Computing and Imaging Institute, University of Utah.\protect
E-mail: \{zhimin, pascucci\}@sci.utah.edu

\IEEEcompsocthanksitem Xin~Yu and Jie~Cao are with School of Computing, University of Utah. E-mail: \{xiny, jcao\}@cs.utah.edu

\IEEEcompsocthanksitem Shusen~Liu, Kailkhura~Bhavya, Diffenderfer~James~Daniel, and Peer-Timo~Bremer are with  Lawrence Livermore National Laboratory. E-mail: \{liu42, kailkhura1,diffenderfer2, bremer5\}@llnl.gov}
}
\IEEEtitleabstractindextext{
\begin{abstract}
Deep learning approaches have provided state-of-the-art performance in many applications by relying on large and overparameterized neural networks.
However, such networks have been shown to be very brittle and are difficult to deploy on resource-limited platforms.
Model pruning, i.e., reducing the size of the network, is a widely adopted strategy that can lead to a more robust and compact model.
Many heuristics exist for model pruning, but our understanding of the pruning process remains limited due to the black-box nature of a neural network model.
Empirical studies show that some heuristics improve performance whereas others can make models more brittle or have other side effects. 
This work aims to shed light on how different pruning methods alter the network's internal feature representation and the corresponding impact on model performance. 
To facilitate a comprehensive comparison and characterization of the high-dimensional model feature space, we introduce a visual geometric analysis of feature representations.
We decomposed and evaluated a set of critical geometric concepts from the common adopted classification loss, and used them to design a visualization system to compare and highlight the impact of pruning on model performance and feature representation.
The proposed tool provides an environment for in-depth comparison of pruning methods and a comprehensive understanding of how model response to common data corruption.
\major{By leveraging the proposed visualization, machine learning researchers can reveal the similarities between pruning methods and redundant in robustness evaluation benchmarks, obtain geometric insights about the differences between pruned models that achieve superior robustness performance, and identify samples that are robust or fragile to model pruning and common data corruption.}
\end{abstract}
\begin{IEEEkeywords}
neural network pruning, robustness, XAI, information visualization
\end{IEEEkeywords}

}

\maketitle

%%%%%%%%%%%%%%%%%%%%%%%%%Teaser image %%%%%%%%%%
%\begin{figure*}[t]
%\includegraphics[width=\linewidth]{image/Tesear.pdf}
%\centering
%\vspace{-6mm}
%\caption{View (1) on the left shows a fault tolerance boundary visualization of the conjugate gradient algorithm. An execution interval is selected in the boundary view, and the related bit-flip outcome over each bit in the interval is displayed in (3). The error propagation that starts from the selected interval is highlighted in view (2), and similar error propagations are  clustered together. Views (4),(5), and (6) on the right coordinate with each other to demonstrate a nonmonotonic error propagation case in which a large error is injected in the middle of the program but mitigated.
%}
%\vspace{-3mm}
%\label{fig:teaser}
%\end{figure*}

\IEEEdisplaynontitleabstractindextext

% \fi
\IEEEpeerreviewmaketitle

\section{Introduction}

\IEEEPARstart{R}{ecent} developments in deep learning have produced significant advances in a variety of application areas~\cite{silver2016mastering, jumper2021highly,senior2020improved}. 
However, such performance is often achieved through extremely large neural networks that consume substantial resources.
Further, these models are difficult to deploy and prone to overfitting, leading to poor generalization and fragile behavior~\cite{huang2019gpipe}.
Network pruning, which removes neurons and/or connections from a model, is a common approach to mitigate some of these challenges since compressing models can reduce both their computational footprint and their inherent redundancy~\cite{han2015deep,liu2017learning} without significant performance loss.
Although model pruning can be accuracy as the original dense models, some recent works~\cite{hooker2019compressed, liebenwein2021lost} have demonstrated that the resulting sparse models are brittle to out-of-distribution shifts~\cite{bulusu2020anomalous}.
For example, common, real-world corruptions can reduce the accuracy of such models by up to 40\% for images from ImageNet-C~\cite{hendrycks2019benchmarking}.
This degradation of robustness has raised serious concerns about the practical viability of pruned models, especially in safety-critical applications such as autonomous driving. 

Recent results~\cite{diffenderfer2021winning} have demonstrated both theoretically and empirically that these problems are a byproduct of the pruning methodologies rather than a fundamental limitation of sparse networks.
Previous work~\cite{diffenderfer2021winning} has theoretically indicated that sparse networks with accuracy and robustness comparable to dense models exist. 
Furthermore, in some instances, it has been empirically demonstrated that pruning can in fact improve both the accuracy and robustness of models compared to their dense baselines. 
This finding is especially surprising as making any model, let alone a pruned version, more robust to out-of-distribution shift has proven difficult.
Nevertheless, it remains unclear why certain pruning techniques positively or negatively affect robustness. 
Providing an in-depth understanding will not only support a real-world deployment of such models but  also might lead to even more advanced pruning approaches.
%Recently, the authors in~\cite{diffenderfer2021winning} made a breakthrough showing, both theoretically and empirically, that these negative results are a byproduct of inapt pruning strategies, and the inability to create pruned models is that perform similar to dense model in a broad range of metrics not fundamental. A somewhat counter-intuitive finding from \emph{``A Winning Hand"} paper is that in sharp contrast to the popular belief that pruning is harmful to the robustness, they show that a certain class of pruning methods (referred to as ``Lottery Tickets") are capable of providing significantly improved robustness over their dense counterparts. In fact, models that are more accurate on the out-of-distribution data relative to the dense baseline are exceedingly rare. 
%\emph{Understanding why and when some pruned models exhibit corruption robustness is key to enabling real-world deployment of machine learning (ML).}
To date, it is unclear what properties of these models can be attributed to their improved performances, and the model pruning community does not have comprehensive introspection tools to answer these important questions. 
Such an effort is hampered by the opaque nature of neural networks and the lack of a dedicated system for model comparison and evaluation in the context of neural network pruning. 

In this paper, we aim to fill this crucial gap by introducing a visual analytical system for understanding differences among representative pruning methods and measuring and interpreting model behavior under various pruning strategies. 
As a general goal, we hope to understand the effect of model pruning on multiple levels, e.g., why some samples are more affected by pruning\cite{hooker2019compressed}, why certain pruned models~\cite{diffenderfer2021multiprize} can have better generalization performance than a state-of-the-art dense-weight trained model, and how the performance of pruned models differs for unseen or corrupted data, etc.
To achieve the goal of building a comprehensive introspection system for analyzing model pruning, our tool focuses on both the computation and visualization fronts.

On the computation side, one essential challenge arises from the need to compare the latent representations of models to understand how making changes to them affects the final prediction. 
However, a neural network's feature representation usually lies in a high-dimensional space that contains hundreds if not thousands of dimensions without explicit semantics or labels.
Comparing such spaces is a nontrivial task, especially considering the behavior of a classifier can be sensitive to small changes (e.g., adversarial example) in the feature representation. 
Traditional dimensionality reduction methods \cite{van2008visualizing, wold1987principal} are not suitable solutions because, for complex latent spaces, they will invariably induce information loss that could significantly impact the trustworthiness of the downstream analysis.
%this can move the detailed discussion 
% Here, we focus on the analysis of the feature representation layer before the final task layer, i.e., a classifier, as it captures high level semantics and encapsulates all the changes that will affect the final prediction.
A potential solution to this challenge is to preserve high-dimensional relationships in the data for our comparison task. 
Since our goal is to understand how latent space changes affect the final prediction, e.g., image classification result, what aspects of the feature representation directly contribute to the prediction is critical. 
As long as this information is encoded faithfully, then the comparison of the high-dimensional feature representations can be more meaningful.

We propose a set of geometrically inspired features (namely \emph{Angle}, \emph{L2 Norm}, \emph{Margin}) derived from a direct decomposition of the classification loss function (i.e., cross-entropy) to evaluate pruning and data corruption. 
These geometric features capture aspects of the feature representation that are directly linked to the model's prediction, which allows us to achieve a comparison of network representations by isolating the most crucial information while removing other variations and noises.
Moreover, since many crucial insights can be obtained only through a comparison between different methods, the overall design of the linked interfaces is centered around the ability to provide a contrastive visualization.% i.e., visualize the difference, or provide the same view of the two datasets side by side.

\major{Based on the analytical framework, we design a novel visualization system which supports comparison with three level of details: pruning methods and evaluation benchmark comparison; feature representations' geometric comparison; detail sample's input feature comparison.
%reveals the difference between pruning solutions with local, and global geometric comparison, and geometric feature attribution.
By utilizing the proposed visual analytic system, researchers can compare pruning methods, reveal the similarities of robustness evaluations benchmark, understand where and how pruning methods differ, identify if a subset of samples is vulnerable to model pruning and data perturbation, and provide insights into why one model is more robust than another. 
These observations can provide feedback to streamline our analysis, improve our understanding of neural network pruning, and motivate useful hypotheses for domain experts to further improve a model's performance}.

Our key contributions are summarized as follows:
\begin{itemize}
\item A geometric description of the structure of a neural network model's feature representation with three geometric metrics to evaluate and compare model pruning and data corruption techniques (section 5).

\item \major{A new dedicated introspective visualization system with a three-level hierarchical comparison based on geometric features for analyzing and comparing major pruning methods over different architectures, corruption datasets, and samples (sections 4 and 6)}.

%\item \major{A novel and scalable visual solution that leverages these geometric metrics to compare the geometry of hundreds of models' feature representation(section 6.3).}

\item Extensive use case that involve state-of-the-art models to demonstrate the usability of the proposed visualization system for pruning and robustness analysis (section 7).
\end{itemize}

%We evaluate the usability of these three metrics comparing the geometry structure to understand different models with different performance. Give the information that critical for the classification.
%%For design such a visualization, we use the semantic direction in the neural network model for each classification category and use model's built-in weights to construct the semantic direction.
%With the semantic direction, we introduce two global features, confident metric and discrimination metric, to explain how a model process latent feature embeddings with these two features.
%We evaluated the usability of three metrics over model pruning, data perturbation and attribution base interpretability and demonstrate the usability of the system over state of art robustness model.

%In this work, we introduce a new visualization system to evaluate the effect of network pruning. 
%Beside the prediction accuracy, our system enable to compare multiple models with different performance metrics at the same time. 
%We discover a semantic direction build-in in the neural network and introduce three global features to explain how a model process latent feature embedding. Based on the metric, we design a global summary view to demonstrate the dynamic of model over pruning and data perturbation.

% (For writing) Unify language and description
% \begin{itemize}
%     \item geometric descriptor
%     \item 
% \end{itemize}
\section{Related Work}

In this section, we discuss various directions that are related to neural network pruning and visualization approaches and discuss their relationship with respect to the proposed approach. %how the proposed  different information reduction techniques.

\subsection{Network Pruning}
In this work, we focus on evaluating and comparing neural network pruning approaches~\cite{lecun1989optimal, frankle2018lottery, ramanujan2020s, diffenderfer2021multiprize, hooker2019compressed}, most of which originate in the ML community. 
LeCun et al.~\cite{lecun1989optimal} proposed a pruning method based on the assumption that an optimized neural network model can reach a function's minimum, and its second derivative can indicate the importance of weights.
Frankle and Carbin~\cite{frankle2018lottery} proposed a lottery ticket hypothesis that a sparse subnetwork with the same initialization can be as accurate as a dense network after training.
Ramanujan et al.~\cite{ramanujan2020s} and Diffenderfer and Kailkhura \cite{diffenderfer2021multiprize} showed that an untrained subnetwork that has the same performance as a weight-trained model. 
Hooker et al.~\cite{hooker2019compressed} introduced pruning-identified exemplars (PIE), which highlight a subset of samples that are more vulnerable to pruning than the other samples. 
To provide a more comprehensive understanding of these techniques, we discuss the pruning problem in depth and explain the differences among popular pruning methods in Section~\ref{sec:pruning}. 

\major{A notable visualization work in this context is CNNPruner~\cite{li2020cnnpruner}. 
Li et al. designed a visual analytic system that enables users to interactively perform pruning and explore the trade-off between the model accuracy and pruning ratio. However, their motivation and goal arise from the question of how to design a human-in-the-loop interactive pruning system. Instead, we aim to evaluate and understand different pruning methods and their robustness in relation to a model's internal representation.
Particularly, our framework provides a geometric similarity comparison between pruning methods and geometric insights into samples that are more vulnerable/robust to the network pruning. Our system finds that the random untrained subnetworks that are surprisingly robust to common corruptions have a significant geometry shape compared with regular well-trained models.}

\subsection{Compression/Pruning for Model Explanation}
In the visualization community, apart from the aforementioned interactive network pruning work \cite{li2020cnnpruner}, the majority of related works on network pruning have focused on the model interpretation problem.
Wang et al.~\cite{wang2019deepvid} used model distillation techniques to compress the size of the model and combined it with a deep generative model to understand the model's reaction to a sample's neighbor. 
Summit~\cite{hohman2019s} proposed two aggregation techniques (activation and neuron-influence aggregation) to select critical neurons (e.g., 7.5\% weights) to build the attribution graph. 
With its feature visualization, the system tries to show an overview of the network model. However, the selected attribute graph is a subnetwork, and the accuracy of the model is not verified. 
Liu et al.~\cite{8802509} designed a visualization system to understand how adversarial examples affect a model's prediction by visualizing the critical subnetwork that preserves the same prediction accuracy of the complete network with these selected samples. 
The critical subnetwork is auto-selected by an optimization process. 
Kahng et al.~\cite{kahng2017cti} designed a visualization system, ACTIVIS that is used for a model's hidden layer activation pattern exploration. 
The system designs an activation matrix that shows the top n-th activation neurons one at a time to compare the activation of different samples or the input of different categories. 
The authors used these most active neurons to select the subnetwork. 
% Dimension reduction~\cite{pezzotti2017deepeyes, rauber2016visualizing} is a common solution to visualize the neural network's latent feature representation. It projects samples' high dimensional latent features into 2D space for users to study the relationship of samples. It can also project network neurons' feature into the low dimensional space to explore the activation similarity of neurons in the neural network.
%
Rather than explaining the prediction or focus rationale behind individual prediction, our study focuses on a global comparison of state-of-the-art model pruning methods in the context of model robustness under common corruptions.

\subsection{Model Evaluation and Model Robustness}
Neural networks have shown superhuman performance on clean test datasets but they fall short on robustness by performing poorer on out-of-distribution data~\cite{bulusu2020anomalous}. This brittleness issue is more prominent for pruned models~\cite{hooker2019compressed}, which makes it crucial to evaluate them on common corruptions arising in real-world applications. 
To evaluate the performance on neural networks in the real world, several corruption benchmark datasets have been proposed. 
Hendrycks and Dietterich \cite{hendrycks2019benchmarking} developed corruption robustness benchmarking datasets CIFAR-10/100-C, ImageNet-C, and ImageNet-R to facilitate robustness evaluations of CIFAR and ImageNet classification models. 
Sun et al. \cite{sun2021certified} and Mintun et al. \cite{mintun2021interaction} further designed new corruption types to complement \cite{hendrycks2019benchmarking}. In addition to image classification, benchmarking datasets for object detection and point cloud classification were developed in \cite{michaelis2019benchmarking} and \cite{sun2022benchmarking}, respectively. 
Motivated by the work on corruption robustness benchmarking, in this work we evaluate a range of pruned classifiers not only on clean test data but also on corrupted datasets.

\subsection{Model Comparison}
We perform a short literature review of different visual model comparison approaches.
The common approach compares the input and output of a model to infer its property.
Square~\cite{ren2016squares} designs a visualization of a model's output to compare models' multilabel prediction behaviors.
Manifold~\cite{zhang2018manifold} uses input and output to compare multiple model behaviors, and the comparison may not be constrained by different model architectures or algorithms.
ConfusionFlow~\cite{hinterreiter2020confusionflow} deploys a confusion matrix with a temporal visual encoding that enables users to track class-level temporal information during model training and comparison.
StackGenVis~\cite{chatzimparmpas2020stackgenvis} utilizes multiple output metrics to compare models' performance and available information to assemble more powerful models. Many techniques~\cite{wang2022learning,li2020visual} use input and output analysis to perform the model comparison.

Information extraction from the output of a model is valuable. It provides a confident score of the model's decisions and reveals the ambiguity of samples among multiple categories. However, this information is limited concerning the stability of the prediction (e.g., adversarial example). 
Also, recent research~\cite{guo2017calibration} has found that output probability can be problematic, and a model may be uncalibrated.
Instead of comparing model output, we study the feature representation extracted by a CNN model before classification.
By studying the feature representation of a neural network directly, we are able to have a global intuition about the prediction behavior of a model and gain more information about a model's behavior such as the robustness of a prediction.
Previous studies have often used dimension reduction that projects high-dimensional data into 2D space to study the data cluster~\cite{xia2021revisiting} and sample density or outliers. 
However, knowledge learned from the 2D projected space contains uncertainty~\cite{liu2014distortion} because the projection process may lose a significant amount of information in the original high-dimensional space.
How to properly understand and use the projected 2D space from high-dimensional space is still ongoing research.

In convolution neural networks, the last classification layer is a linear layer, and some of these models are attached to a softmax operation.
Therefore, the last layer's feature representation is more accessible and interpretable than the previous latent space.
Limited work has been designed to understand the latent feature space because of the unknown mysterious structure of the feature latent representation.
In this work, we leverage the loss function to construct global geometric features to understand and compare the behavior of multiple pruning methods.

\section{Domain Background}
In this section, we discuss the basic terminologies, pruning techniques, and evaluation methods used in this study.

\subsection{Network Pruning} 
\label{sec:pruning}
Here, we introduce the pruning approaches used in this study for evaluation and analysis.
In this paper, we will focus on unstructured pruning, which removes redundant network weights. 
Nevertheless, the analysis pipeline also works for structured pruning, which prunes entire neurons or filters. 
%For a model which will be pruned $p\%$ ($p \in [0,100]$) of weights, most pruning approaches will use a heuristic value to rank the importance of the weights in the neural network model and prune the least important weights.% In this study, we use three pruning methods for the model compression. 
As summarized in ~\cite{blalock2020state}, most works in network pruning start with scoring the model parameters based on their potential impact on the network performance, selecting weights of least importance to remove from the network, and optionally performing retraining to gain back performance degradation due to pruning. We will explore the following pruning methods:  

\textbf{Random Pruning}: Randomly select a set of weights and remove them from a neural network model.

\textbf{Magnitude Pruning}: Score the weights with their absolute values and prune the ones with the smallest scores~\cite{han2015deep}.

\textbf{Gradient-Based Pruning}:
Prune the weights that would have the least impact on the loss function of the model if removed. Given the training data $\mathcal{D}$ and the neural network with its weights $\mathcal{\phi}$, the impact on the loss function $\mathcal{L}$  after removing a single weight $\bm{\phi}_m$ can be  estimated as $\mathcal{I}_m$ by the functional Taylor expansion  as 
\begin{align*}
\mathcal{I}_m = |& \nabla \mathcal{L}(\mathcal{D}, \bm{\phi}) (\hat{\bm{\phi}} - \bm{\phi}) \\ & + \frac{1}{2}(\hat{\bm{\phi}}-\bm{\phi})^T \nabla^2 \mathcal{L} (\mathcal{D}, \bm{\phi}) (\hat{\bm{\phi}} - \bm{\phi}) \\ & + O(\|\hat{\bm{\phi}} - \bm{\phi} \|^3)|
\end{align*}
where $\nabla \mathcal{L}$ and $\nabla^2\mathcal{L}$ indicate the first-order gradient of the weights and second-order gradient (Hessian matrix) respectively~\cite{lee2018snip, molchanov2019importance,hassibi1992second}.

\textbf{Multi-prize lottery tickets (MPTs)}: This strategy searches for a performant sparse subnetwork within a randomly initialized network and can further compress the network by applying weight binarization. 
Counter to the traditional training paradigm of learning the network weights, this approach learns which randomly initialized weights should be retained to improve performance by optimizing over surrogate scores that indicate the importance of each weight to network performance. 
In our experiments, we make use of biprop (Algorithm 1 in \cite{diffenderfer2021multiprize}). 
%Motivated by the \emph{strong lottery ticket hypothesis} \cite{ramanujan2020s}, 
This methodology is built on the %even stronger 
\emph{multi-prize lottery ticket hypothesis} \cite{diffenderfer2021multiprize}, which proposes that sufficiently overparameterized randomly initialized networks contain sparse subnetworks that, without any training, can perform comparably to dense networks and are amenable to weight binarization. 
Theoretical proofs supporting this hypothesis have been established \cite{diffenderfer2021multiprize,sreenivasan2021finding}. %,emanuele2022proving}.
Further experimental efforts have established that these MPTs learned using biprop are comparably or more robust than dense networks learned using traditional weight optimization in work demonstrating that certain model compression algorithms are capable of producing compact, accurate, and robust deep neural networks, or CARDs \cite{diffenderfer2021winning}.

%The classical Optimal Brain Surgeon (OBS)~\cite{hassibi1992second} method calculate the importance score with the Hessian matrix and update the remaining weights after pruning $\bm{\phi}_m$ to accommodate its loss. Due to the heavy computation for Hessian matrix, 
%Recent works revisit this gradient-based method and assume that (i) the higher than the first order terms are ignored for efficient approximation (ii) removing $\bm{\phi}_m$ will not affect the remaining weights, so $\hat{\bm{\phi}}_m=0, \hat{\bm{\phi}}_i = \bm{\phi}_i ~\forall i \neq m$. Hence, 
%\begin{align*}
%\mathcal{I}_m \approx  |& \nabla \mathcal{L}%(\mathcal{D}, \bm{\phi}_m) \bm{\phi}_m|
%\end{align*}
%Although we focus on the above method in our paper, our proposed pipeline also works for the other methods derived from OBS~\cite{} with the higher order Taylor expansion.

\subsection{Data Corruption Evaluation}
Besides accuracy on clean test data, we use the cifar10-C dataset~\cite{hendrycks2019benchmarking}, which is the cifar10 test dataset corrupted with 19 common corruption \textbf{\textit{algorithms}} from four categories: noise, blur, weather, and digital corruptions. 
For an image dataset, similar corruption operations can be performed to generate the same corrupted image datasets. For example, the same corruptions have been performed to generate
MNIST-C, Cifar100-C, and Imagenet-C datasets. We also perform a quantitative evaluation of these datasets based on exploration observation.
These corruptions preserve the semantic content of images, and humans can easily recognize these images.
Each corruption technique has a severity level from one to five where a larger number denotes a more severe corruption.

%We define a \textbf{robustness} metric to measure a samples' invariant to different data corruption.
%It is defined as the average accuracy on corrupted data cifar10c (i.e., applying $c=19$ corruption types and $s= 5$ severity levels to each clean test image $x_i$ and transforming it to a set of corrupted images $x_i^{c,s}$):

%\begin{align}
%    robustness = \sum_{i=1}^N robustness_i,\\
%    robustness_i = \sum_{c=1}^{19} \sum_{s=1}^5 \mathcal{I}_{\{y_i = f(x_i^{c,s})\}},
%\end{align}
%where $\mathcal{I}$ is an indicator function taking value 1 if the prediction is correct on corrupted image and 0 otherwise.
%Furthermore, $robustness_i$ denotes per sample robustness quantifying the level of invariant to corruptions. 
\section{Tasks Analysis}\label{sec:task}
As previously mentioned, existing model pruning schemes exhibit diverse behaviors. Specifically, some pruned models are more robust to different data corruption than others. 
Unfortunately, it is not clear why certain pruning techniques affect a model in one way and not another.
In this work, we aim to provide an in-depth understanding of this phenomenon that will not only support a real-world deployment of such models but could also lead to more advanced pruning approaches. Here, we are specifically interested in answering the following questions: For given two models pruned with different pruning techniques, what properties of the models make their accuracy and robustness differ from each other? What properties can make certain samples more vulnerable to the model pruning?
Are current evaluation benchmarks sufficient for the network pruning methods?

These questions, which are critical for comparing and designing better pruning strategies, are articulated by domain experts.
To help domain experts improve their understanding of model behavior and answer the above questions, we distill these questions into the following requirements to drive the design of the visualization system.
The requirements of this work are based on a long-term regular interview  (twice a month over 7 months) with four domain experts who are machine learning researchers and also the coauthors of this work .
\major{Currently, our visualization system is designed for developers who have a machine learning background.}

%Model pruning can be performed with different techniques, and their pros and cons are often evaluated with optimization loss or accuracy. 
%However, this process often misses critical detail of model's inner state dynamic to give an explanation. 

%\BK{Given two models (pruned with different pruning techniques or augmentations), can we explain what aspects of the models make their accuracy and robustness differ from each other? For example, what causes a model to be significantly more robust than the other? Which model has disentangled and better/robust representations for clean vs corrupted data?}

\textbf{R1 - Model Evaluation Overview}: Model pruning experts often perform experiments on multiple architectures, different pruning methods, and various data corruption evaluation benchmarks. A succinct presentation of these results can guide domain experts to figure out the pros and cons of different pruning solutions and architectures and narrow down their analysis to the most interesting subset for a detailed examination.

\textbf{R2 - Evaluation Similarity Between Different Pruning and Corruption Operation}:
Understanding the similarity between pruning and corruption operations will  help domain experts quickly figure out whether one approach is different from another. 
The difference will lead to further examination to understand what makes the two approaches different from each other.
%A pruning profile indicates the amount of weights each layer and each filter are pruned from a neural network model. A redundancy profile of a model reveals the feature redundancy of each neural network layer and helps domain experts to assess whether a pruning strategy is reasonable.}

\textbf{R3 - Model Behavior Over Samples with Different Pruning and Corruption Configurations}: A summary of a model's behavior over different subsets of samples is useful for understanding and diagnosing a model's behaviors under different pruning configurations.   
Meanwhile, a summary of certain representative model properties over a large number of samples is useful for domain experts to diagnose a model's behavior. It helps domain experts understand what mistakes the model will make and what decisions the model will be confident about.
    
    %\item \update{\textbf{R3 - A  Model's Pruning Profile Examination}:}
    %\update{Understand the amount of 
    %A pruning profile indicates the amount of weights each layer and each filter are pruned from a neural network model. A redundancy profile of a model reveals the feature redundancy of each neural network layer and helps domain experts to assess whether a pruning strategy is reasonable.}
    
\textbf{R4 - Latent Feature Representation Variation}: Comparing different models' latent space provides domain experts with a visual understanding of how models' feature representation changes after pruning. Latent feature comparisons are more sensitive than input and output comparisons, which can capture minute changes. Such a comparison can provide valuable insights to understand why a model may be more resilient to input perturbation than another and why certain samples are vulnerable to pruning or corruption.
    
\textbf{R5 - Input Feature's Sensitivity With Respect to Model Pruning and Data Corruption}:
Model pruning leads to varying impacts on a model's decisions on different samples or labels. 
Certain samples are robust to pruning but others are fragile. 
Understanding what input features make a sample vulnerable or robust to network pruning is a critical task for domain experts.

%We formalize the analytical pipeline in Fig.~\ref{fig:analysis_workflow}, and highlight steps that are the main focus of this system.
%\begin{figure}[t]
%\centering
%    \includegraphics[width=\linewidth]{images/5_workflow.pdf}
%    \vspace{-3mm}
%    \caption{The analysis workflow applied in the current visualization system .}
%\label{fig:analysis_workflow}
%\vspace{-2mm}
%\end{figure}
\section{Geometric View of Latent Space}
\label{sec:method}
Fulfilling the above-mentioned requirements is a nontrivial task.
In particular, how to display a model's behavior summary over a large amount of data besides prediction accuracy and how to perform latent space comparisons are difficult questions to answer. %do not have straight forward and intuitive solutions. 
To address these challenges, in this section, we define the class direction and corresponding three geometric metrics (\emph{angle}, \emph{length}, and \emph{margin}) on a neural network's last (a network's layer before classification) layer's feature encoding, which is the accumulated result of previous layers' transformations and directly used for a model's final prediction.

\begin{figure}[t]
\centering
\vspace{-2mm}
    \includegraphics[width=\linewidth]{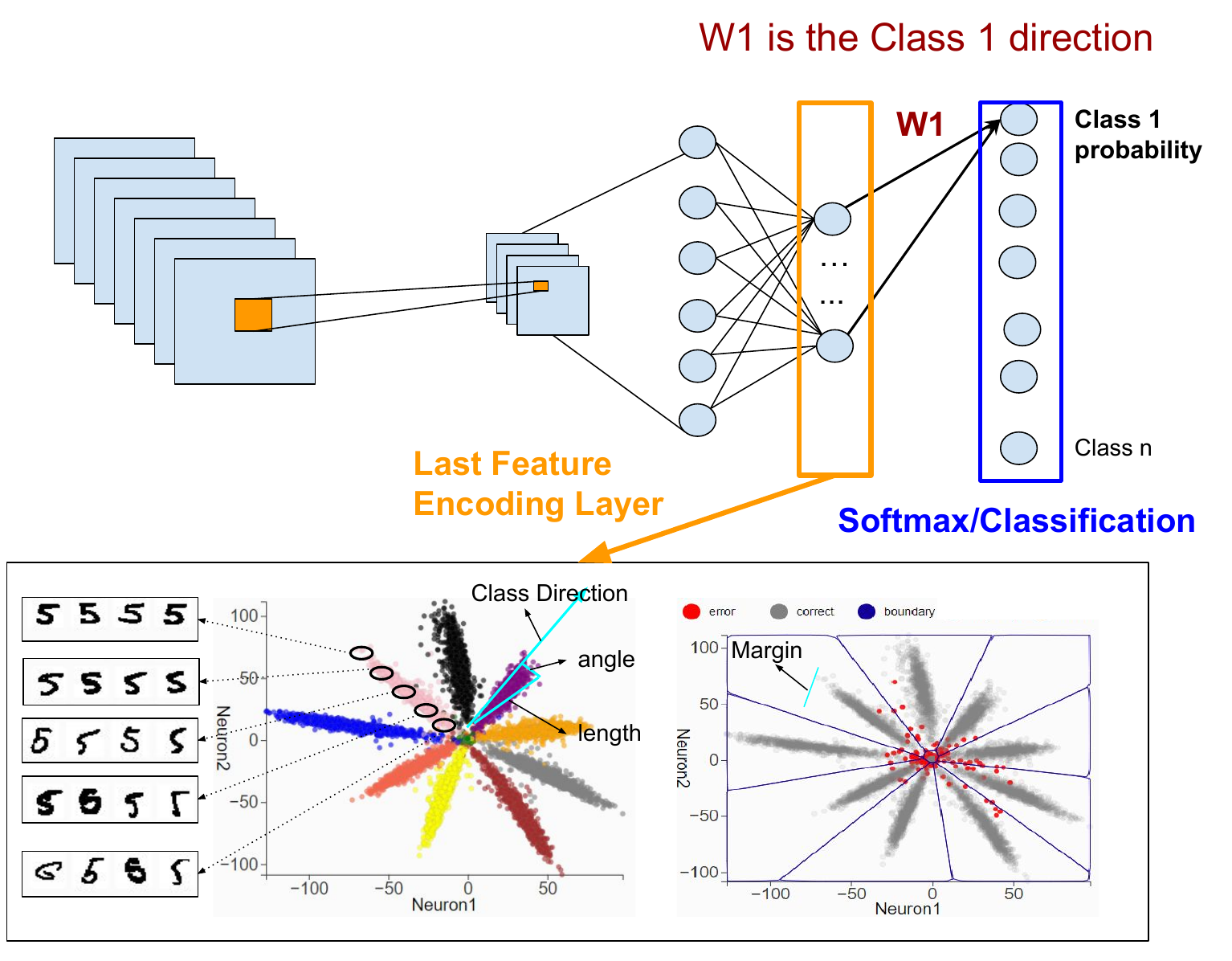}
   
    \vspace{-3mm}
    \caption{This study monitors the last feature encoding a layer's latent feature representation to understand the impact of network pruning.
    The feature representation of data samples in the last feature encoding layer with 2 neurons can be visualized directly. The star shape is the sample's latent space representation of a model trained with the cross entropy loss. The decision boundary in the layer can also be approximated and visualized by sampling the 2D space. Four geometric features that are connected with the loss function can be identified in the visualization.}
\label{fig:a_simple_demo_of_geometric_feature}
\end{figure}

Before getting into a detailed description of the geometric features, we present an example to explain the intuition of what the geometric features look like in the last feature encoding layer.
Fig.~\ref{fig:a_simple_demo_of_geometric_feature} highlights the critical concepts in the following discussion. The figure visualizes the feature representation of samples in a trained neural network, where the last feature encoding layer is set with only 2 neurons. 
These 2D latent feature vectors display a star shape, in which samples that are far away from the origin have more distinguishable features, and samples that are close to the origin are ambiguous (refer to support material for more discussion about the geometric shape and loss function).
Class direction is the direction across the middle of samples that belong to a certain category.
\emph{Angle} and \emph{length} metrics are marked in the plot, which is the angle with the class direction and a sample's distance to the origin, respectively. 
The plot on the right shows the same feature vectors but with the decision boundaries of the classification.
These geometric properties not only exist in 2D space but also can be generalized to high-dimensional space to help summarize crucial aspects of the latent space structure and form the basis for cross-model comparison. 
%At the end of this section, we will discuss the impact of the curse of dimensionality and also how these geometric metrics' correlation with model robustness.

\subsection{Class Direction}
The loss function used for training the neural network is critical for shaping the geometry of the latent feature embedding.
Cross-entropy loss and negative log-likelihood loss are default loss functions used for classification tasks. 
%In this paper, we consider the general hard-classification problem, where an input example can have one and only one category.
For a given example $x$ with a ground truth label $y=i$, the loss function can be formulated as $L_{loss} = -log(P(y=i|x))$ where $P(y=i|x)$ is the predicted probability for a model for the label $y=i$ with a value range in $[0, 1]$. A large $P(y=i|x)$ will achieve a small loss $L_{loss}$.

Without loss of generality, we assume that the neural network is composed of two parts: an encoder $h$ that transforms input $x$ into a feature representation vector $\vec{X}=h(x)$ and a classifier $c$ that is used for producing the predicted probability $P(y=i|x)=c(\vec{X})$. 
We consider that the classifier part $c$ consists of a fully connected layer, and a soft-max activation function, which is a general configuration in convolutional neural network models. 
$\mathcal{C}$ is the number of classes for a classification task and $\vec{X}$ is the $m$-dimensional feature embedding of $\vec{X} \in \mathbb{R}^m$. The last fully connected layer in classifier part $c$ parameterized with weight $\mW \in \mathbb{R}^{m \times \mathcal{C}}$ will take this $m$-dimensional feature vector as input and project it into $\mathcal{C}$ scores, and then output the predicted probability via the soft-max activation function. 
These operations can be summarized as equation~(\ref{eq:0}).
%%%%%%%%%%%%%%%%%%%
%\XY{-----------}
%Let $\vec{X} \in \mathbb{R}^m$ be the feature embedding input for the last layer of the classifiers in a well-trained model. $\mathcal{C}$ is the number of classes for the classification task. 
%We consider that the classifier part consists of a fully-connected layer, and a soft-max activation function, which is a general configuration in CNNs. The last fully-connected layer parameterized with weight $\mW \in \mathbb{R}^{m \times \mathcal{C}}$ will take the feature vector $\vec{X}$ as input, and project it into $\mathcal{C}$ scores, then output the predicted probability via soft-max activation function. 
%\XY{-----------}
%%%%%%%%%%%%%%%%%%%%

In equation~(\ref{eq:0}), we can rewrite the weight matrix $\mW = \{\vec{W}_i|0 < i \leq \mathcal{C}, \vec{W}_i \in \mathbb{R}^m\}$ as the set of weights corresponding to each of the $\mathcal{C}$ classes. 
Hence, the predicted probability $P(y=i|x)$ is determined by the dot products of feature representation $\vec{X}$ and each target label $j$'s neuron weight $\vec{W}_j$.

\begin{align}
P(y=i|x) & = \frac{e^{\vec{W}_i \cdot \vec{X} }}{\sum^{\mathcal{C}}_{j=0} e^{\vec{W}_j \cdot \vec{X}}} =
\frac{e^{\|\vec{W}_i\| \| \vec{X} \| \cos\theta_i}}{\sum^{\mathcal{C}}_{j=0} e^{\|\vec{W}_j\| \|\vec{X}\|  \cos\theta_j}} \label{eq:0}  \\ 
& = \frac{1}{\sum^{\mathcal{C}}_{j\neq i} \frac{e^{\|\vec{W}_j\| \|\vec{X}\|   \cos\theta_j}}{e^{\|\vec{W}_i\|  \|\vec{X}\| \cos\theta_i}} + 1}  \label{eq:2} \\
& = \frac{1}{\sum^{\mathcal{C}}_{j\neq i} e^{\|\vec{X}\| (\|\vec{W}_j\| \cos\theta_j - \|\vec{W}_i\|   \cos\theta_i)} + 1}  \label{eq:3} \\
& = \frac{1}{\sum^{\mathcal{C}}_{j\neq i} e^{\|\vec{X}\|(C_j  \cos\theta_j - C_i  \cos\theta_i)} + 1} \label{eq:4}
\end{align}

The dot product operation (denoted as $\cdot$) can be interpreted as the feature embedding $\vec{X}$ of every input example being projected onto each label $j$'s $\vec{W}_j$ direction and multiplied with $\|\vec{W}_j\|$.
Here, $\|\vec{W}_j\|$ is a constant, and the predicted probability is determined by the L2 norm $\|\vec{X}\|$ and the angles $\theta_{j}$ between the directions of $\vec{X}$ and each $\vec{W}_j$. 
Based on the above intuition, we define the fixed weight vector $\vec{W}_j$ parameterizing the classifier part $c$ as the \textbf{class direction} of category $j$ in the network's last layer latent space. 

%During the training, the optimization process will jointly optimize the feature representation $\vec{X}$ and the $\mathcal{C}$ class directions via propagating the loss backwards to the encoder $h$ and the classifier part $c$.
%Because we assume the classifier part $c$ is fixed after training, the pruning task is to find a tailored encoder part $h^{-}$. After pruning, for each input example $x$ with target label $i$, $h^{-}$ can still project $\vec{X}$ into $\vec{W_i}$ direction with small $\theta_{i}$ and make the L2 norm as large as possible to maintain the correct prediction.

%\begin{figure}[t]
%\centering
%    \includegraphics[width=\linewidth]{images/4_2d_distribution.pdf}
%    \vspace{-3mm}
%    \caption{2 neuron layer in neural network}
%\label{fig:2d_distribution}
%\vspace{-2mm}
%\end{figure}

%During the training, neural network will adjust the weight of the model and make last layers' feature embedding's angle to the $\vec{W_i}$ as small as possible and increase its L2 norm $\|X\|$. Training with the cross-entropy loss, the neural networks' final layer's embedding will coverage to a star shape.  

%The $L2$ norm lead to an unexpected result of the model training which make model over-confident ~\cite{guo2017calibration} with their prediction. 

%\subsection{Equal Separate Space}
%Previous section has define the $\mW_i$ as the semantic direction of the category $i$. One question is the role of scale value of $\|W\|$. The weight $\|W_i\|$ if significant large than $\|W_j\|, j\neq i$ can cause problem.
%$\|W_i\| \approx \|W_j\|$
\subsection{Angle, Length, and Margin}
Previous discussion has defined the class direction that can be extracted from the built weights in the neural network model.
Based on the above definition, we introduce three geometric features of the last layer latent space - angle, length, and margin.

%, and the impact of data corruption on them. %\JC{? Will we introduce the discrimination, confident, robustness here? or we just introduce the angle, L2, and minimum distance}

\textbf{Length metric} is the L2 norm of feature vectors $\vec{X}$ that mainly affects the confidence or output probability of a prediction.

\textbf{Angle metric} is the geometric angle $\theta$ between class directions and feature vectors. %To understand why angle metric matters, we can examine the softmax function in formula ~(\ref{eq:0}) in which the angle is $\theta_i$.
%When pruning a well-trained neural network with the classifier part $c$ fixed, $\|\vec{W}_i\|$ or $\|\vec{W}_j\|$ is a constant value that does not get affected by the feature embedding $\vec{X}$.
%Since a given prediction shares the same $\|\vec{X}\|$, the angle value $\theta$ between class directions and feature vectors is the only variable that determines the prediction result. 
A small angle between the class direction $\vec{W}_i$ means the sample will be predicted as label $i$ with high probability. 
If a feature vector has similar angles with respect to two class directions, then the model will give similar probability to both categories.
If the model makes an incorrect prediction on a sample, then the feature vector of this sample often has a large angle with the target class direction. 
This metric can be affected by the curse of dimensionality. See support material for more detail.

%Assume the angle between feature vectors and the other class directions maintain the same, decreasing the angle $\theta_i$ will decrease the value of equation~(\ref{eq:5}) which will increase the probability of equation~(\ref{eq:4}). 

\textbf{Margin metric} is the minimum distance to the decision boundary that is often interpreted as a prediction's robustness. A large margin can tolerate severe data corruption and large perturbations in the input sample. 
In equation~(\ref{eq:0}),  a sample with the feature embedding $\vec{X}$ belonging to label $i$ needs to have the largest output probability of the model that needs to satisfy 
$$
\vec{W_i} \cdot \vec{X} - \vec{W_j} \cdot \vec{X} > 0,j\neq i, j \in {1,..,\mathcal{C}}.  
$$

The decision boundary of label $i$ is constructed with $n-1$ hyper-planes $
(\vec{W_i} - \vec{W_j}) \cdot \vec{X} = 0,j\neq i, j \in {1,..,\mathcal{C}} 
$. 
The minimum distance of a feature vector to the decision boundary is the minimum distance of the feature vector $X$ to all $n-1$ hyper-planes:
\begin{equation}\label{eq:5}
min\{\frac{\|(\vec{W_i} - \vec{W_j}) \cdot \vec{X}\|}{\|(\vec{W_i} - \vec{W_j})\|}, \quad j\neq i, \; j\in 1,..,\mathcal{C}\}.
\end{equation}

Note that if a model makes a wrong prediction on a sample, then the margin value will be multiplied by a negative one to indicate the error. %\update{A comparison between three geometric metrics and output probability is presented in the support material.}
%the amount of margin value to be fixed the prediction error.
%One thing can be noticed in equation~(\ref{eq:5}) is  $\|(\vec{W_i} - \vec{W_j}) \cdot \vec{X}\| = \|(\vec{W_i} - \vec{W_j})\|* \|\vec{X}\| * \cos(\theta)$. Assume $\mW$ and $\cos(\theta)$ do not change, increase the l2 norm of the feature embedding will increase the minimum distance to the decision boundary. In alternative, increase the probability of the prediction will increase the robustness of the prediction.

\major{A similar concept is described as visual angle hardness~\cite{chen2020angular}, in which normalized angles and l2 are used as a metric to describe the generalization of neural network prediction.
Liu et. al~\cite{liu2016large} proposed a large-margin softmax loss that was used to maximize the margin between two face recognition predictions.
Compared with previous studies, ours combines these metrics as a whole to describe the geometry of a neural network's high-dimensional feature representation and embeds these metrics into a visualization system to compare the similarity between neural network models and monitor the impact of network pruning and data corruption.}

\section{System Design}

\begin{figure*}[t]
\centering
    \includegraphics[width=\linewidth]{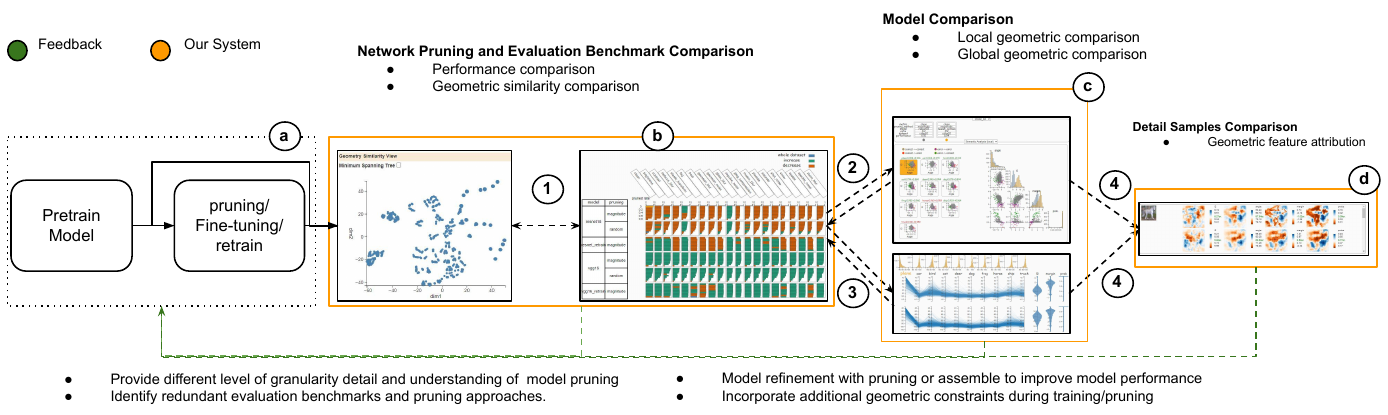}
    \vspace{-5mm}
    \caption{\major{Compared with classical pruning analysis (\textcircled{\small{a}}), the workflow of our visualization system is embedded with three levels of comparison: pruning method and evaluation benchmark comparison; model comparison; sample comparison. 
    Domain experts can compare pruning techniques and data corruption benchmarks (\textcircled{\small{b}}) through the interaction between the evaluation view and geometric similarity view. 
    For selected models (\textcircled{\small{c}}), users can examine and compare their local and global geometric properties. 
    Furthermore, the sample level comparison (\textcircled{\small{d}}) enables users to examine features that are captured by models.    
    During the exploration process, users can choose a pruning technique to refine the original model, remove redundant evaluation benchmarks, assemble multiple pruned models for better performance, or incorporate geometric constraints for model retraining.}}
    %Users can interact (\textcircled{\small{1}}) between evaluation table view and geometric similarity view to compare the similarity of pruning methods and evaluation benchmark. With selected interesting models and benchmarks, user can examine their detail geometric properties (\textcircled{\small{2}}) through local geometric view and global geometric view. In the meantime, user can select a subset of samples(\textcircled{\small{3}}) in local geometric view or global geometric view base on users' interests to examine their performance in different corruption benchmark and pruning methods. At the end, user can examine a detail sample's feature attribution (\textcircled{\small{4}}) to compare what features are captured by different models.}}
\label{fig:workflow}
\vspace{-3mm}
\end{figure*}

\major{Leveraging previously proposed geometric metrics,
we design a visualization system that generally follows the overview first (Fig.~\ref{fig:workflow}(\textcircled{\small{b}}) and detail-on-demand (Fig.~\ref{fig:workflow}\textcircled{\small{c}}, \textcircled{\small{d}}) mantra~\cite{shneiderman2003eyes}.
We introduce five key visual components that integrate these metrics and coordinate with each other for pruning evaluation and model comparison.}

\major{In Fig.~\ref{fig:workflow}, we demonstrate a workflow overview of our visualization system. 
Fig.~\ref{fig:workflow} (\textcircled{\small{a}}) is the classical work pipeline of machine learning engineers for performing model pruning~\cite{liu2018rethinking}. 
Such a pipeline often evaluates the model performance with prediction accuracy but ignores the impact of pruning over a model's robustness, latent space variance, and sample-level impact.
Beyond the classical pruning analysis pipeline, our system provides broader evaluation~\cite{tulio2020beyond}, model comparison, and geometric exploration.
Furthermore, the system also supports fine-grain sample-level analysis from a whole dataset to a single sample.
Such interactions provide domain experts insights about pruning behaviors and give them feedback to improve the evaluation methods and neural network models.
In the following section, we will discuss the design rationale for these visual components and explain how they work together to address the corresponding domain requirements (see section \ref{sec:task}).}

\subsection{Evaluation Table and Geometric Similarity View}
\major{Having an overview of pruned models' performance over different evaluation benchmark and their feature representation similarity is an essential first step (\textbf{R1}).
This process involves comparison among multiple model architectures, pruning approaches, evaluation metrics, and different subsets of samples' prediction outcomes.
However, efficiently comparing the feature representations of hundreds of models visually is a challenge.
To address the above requirement, we design an evaluation table view for performance comparison and a geometric similarity view to compare geometric similarity between models' feature representations.}

%We aggregate this information into a table format and design the evaluation table view and model similarity overview.

\begin{figure}[t]
\centering
    \includegraphics[width=\linewidth]{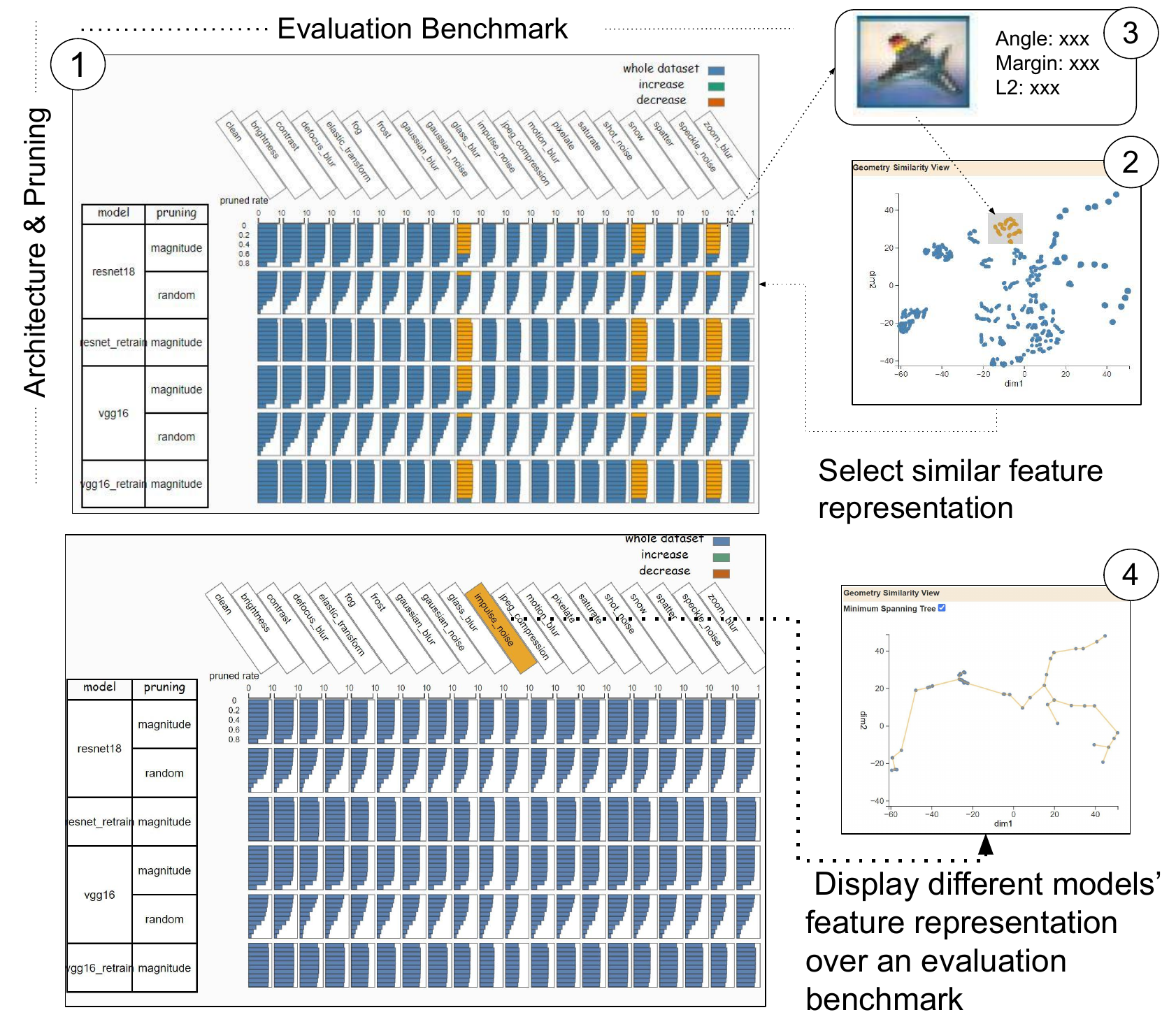}
   
    \vspace{-3mm}
    \caption{\major{The evaluation table view and geometric similarity view can coordinate with each other to scalably compare many models at the same time. Users can select a set of nearby points that are similar models in the geometric similar view\textcircled{\small{2}}, and the related model will be highlighted in the evaluation view\textcircled{\small{1}}. A similar operation can also be performed in the evaluation table view.}}
\label{fig:evaluation_table_and_geometric_similarity}
\vspace{-3mm}
\end{figure}

\major{In Fig.~\ref{fig:evaluation_table_and_geometric_similarity} (\textcircled{\small{1}}), the evaluation table displays an evaluation summary of the models' performance over different data corruption benchmarks~\cite{hendrycks2019benchmarking}. 
The x-axis of the evaluation table represents different data corruption types, and the y-axis represents the architectures and pruning techniques.
In the visualization, each histogram represents the evaluation outcomes (i.e., accuracy) on a given corruption-type dataset, for models with increasing pruned rates, which is defined as the fraction of weights removed by the pruning algorithm. 
With the increasing pruned rates, the performance of the model often worsens except for the model with retraining.}

\major{Comparing the similarities among neural network models is critical for the pruning approach analysis (\textbf{R2}).
In Fig.~\ref{fig:evaluation_table_and_geometric_similarity} (\textcircled{\small{2}}), we design the geometric similarity view to compare the similarity of pruned or retrained models' behavior over different data corruption benchmarks. 
For each model, we use each sample's three geometric features--\emph{angle}, \emph{margin} and \emph{l2} norm (Fig.~\ref{fig:evaluation_table_and_geometric_similarity} \textcircled{\small{3}}) as properties to describe a model.
A model  is evaluated with 10,000 samples will have 30,000 geometric features.
The geometric similarity view projects all models' high-dimensional features into two-dimensional space by the UMAP~\cite{mcinnes2018umap} dimension reduction techniques.
In the visualization, each point represents a model, and nearby points (models) are more similar than the others.}

\major{The evaluation table view and the geometric similarity view can interact with each other.
A set of models that have similar feature representation is selected in geometric similarity view (Fig.~\ref{fig:evaluation_table_and_geometric_similarity} \textcircled{\small{2}}), and related models are highlighted in the evaluation table view (Fig.~\ref{fig:evaluation_table_and_geometric_similarity} \textcircled{\small{1}}).
Similarly, the model selection operation can also be performed in the evaluation table view and relative models will be highlighted in the geometric similar view.
In Fig.~\ref{fig:evaluation_table_and_geometric_similarity} \textcircled{\small{4}}, we select models that are evaluated with the \textit{impulse noise} corruption benchmark. 
The geometric similarity view displays only models that belong to this evaluation.
Furthermore, these points can be connected with a minimum spanning tree for similarity comparison between different evaluation benchmarks (see use case 7.1 for more detail).}

\subsection{Local Geometry View} 
\major{The model evaluation table view and geometric similarity view provide a summary of the overall performance and similarity between network pruning techniques.
Beyond the similarities and differences between network pruning approaches and evaluation methods, it is important to have a detailed geometric examination of each feature representation.}

\begin{figure}[t]
\centering
    \includegraphics[width=\linewidth]{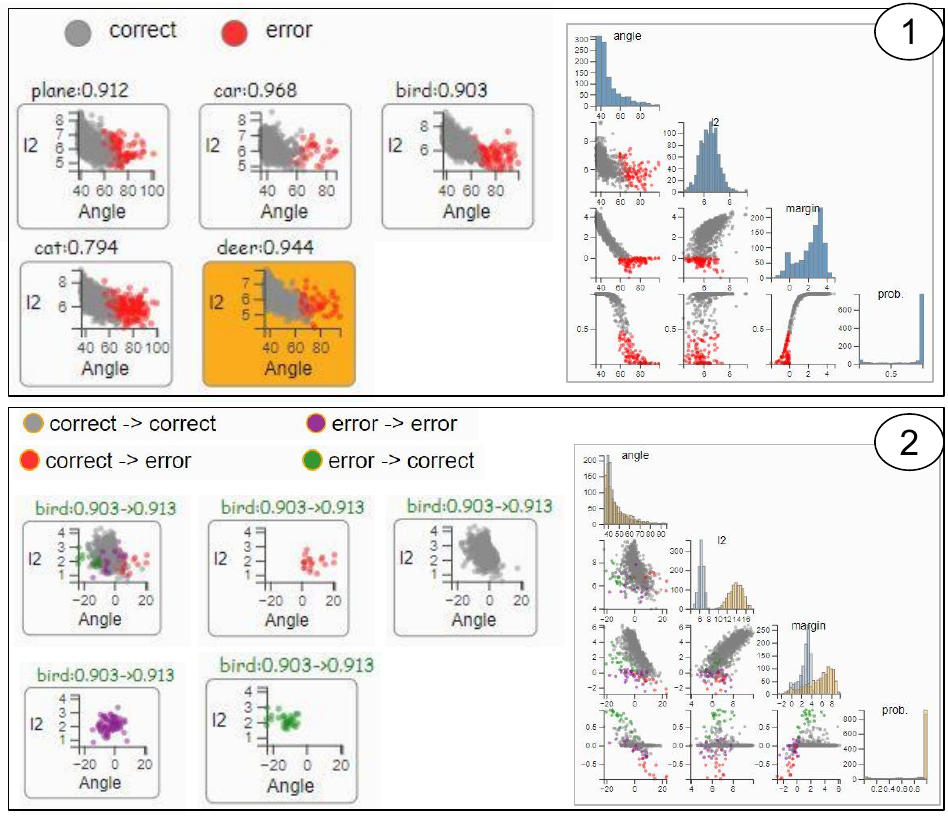}

    \vspace{-3mm}
    \caption{\major{Local geometric view visually presents the geometric metrics distribution of each label category \textcircled{\small{1}}. The incorrect predictions are samples often with large \emph{angle} and small \emph{l2} norm. Furthermore, the local geometric view also enables users to compare two models' geometric differences (\textcircled{\small{2}}) by displaying the geometric value variation between samples from different models and providing highlight operations for samples that are errors in one model but correct in the other.} }
\label{fig:local_analysis_pipe_line}
\vspace{-2mm}
\end{figure}

The Local Geometry Plot performs class-specific evaluations, which helps in conveying how well samples from a specific class are classified and uncovers the cause of potentially poor performance.
The visualization in Fig.~\ref{fig:local_analysis_pipe_line} \textcircled{\small{1}} displays geometric metrics (angle and l2 norm) for five classes, and the deer class is selected to examine more local features. 
The accuracy on the top shows that samples from the cat class have 79.4\% accuracy, which is less than the rest of the samples, and the car class has the best performance with 96.8\% accuracy.
The \emph{angle} and \emph{length} metrics are presented as a scatter plot.
Within a single model, these geometric metric values often have similar ranges for each class.
Among these points, the gray color indicates the correctly classified samples, and the color red indicates the incorrectly classified samples. 
In all five categories, these incorrect samples have larger \emph{angle} and smaller \emph{length} metrics than the correct samples.
By selecting a category, a user can examine more of a class's local features, including \emph{margin}, and \emph{probability}.

\major{In addition to exploring a model's latent space geometry and its per-class performance, the local analysis also facilitates a comparison between models (\textbf{R4}).
In Fig.~\ref{fig:local_analysis_pipe_line} \textcircled{\small{2}}, we select models with pruning and retrain for comparison.
The visualization shows the variation of the geometric metric distribution of bird samples and the corresponding geometric value change.
Users can selectively pick samples that exhibit diverse reactions to network pruning and retraining.
We encode these reactions into four types based on domain experts' recommendations. One of these types is that a sample is predicted correctly in one model but wrong in the other (red). In the opposite case, a sample is predicted wrong in the original model but incorrect in the new model, which is encoded as green.
The value in each axis is the difference (e.g., $\Delta$ angle = angle - angle') of  samples' geometric value between two models. The related scatter plot matrix encodes all geometric feature distributions and their distribution shift.}

\subsection{Global Geometry View}
The local geometry plot provides a detailed view of the behavior of a single class. However, for the comparison task, the ability to have a more comprehensive summary is crucial.
A global geometry plot gives a geometric overview of how a model performs on all classes of the currently selected dataset (\textbf{R2}), which shows not only how well samples are classified, but also what other classes the model is confused with. 

The global geometry plot uses a parallel coordinate display in a $n+3$ dimension configuration, in which $n$ is the number of classes (for a dataset with a large number of classes, a pres-election can be applied to focus on a subset of classes to make visual encoding and exploration manageable). 
\major{The first $n$ dimensions represent the \emph{angle} metric of a sample with respect to each class direction. As the previous section described (section 5), a relatively small angle with a label means a good chance the sample belongs to this category.}

\begin{figure}[t]
\centering
    \includegraphics[width=\linewidth]{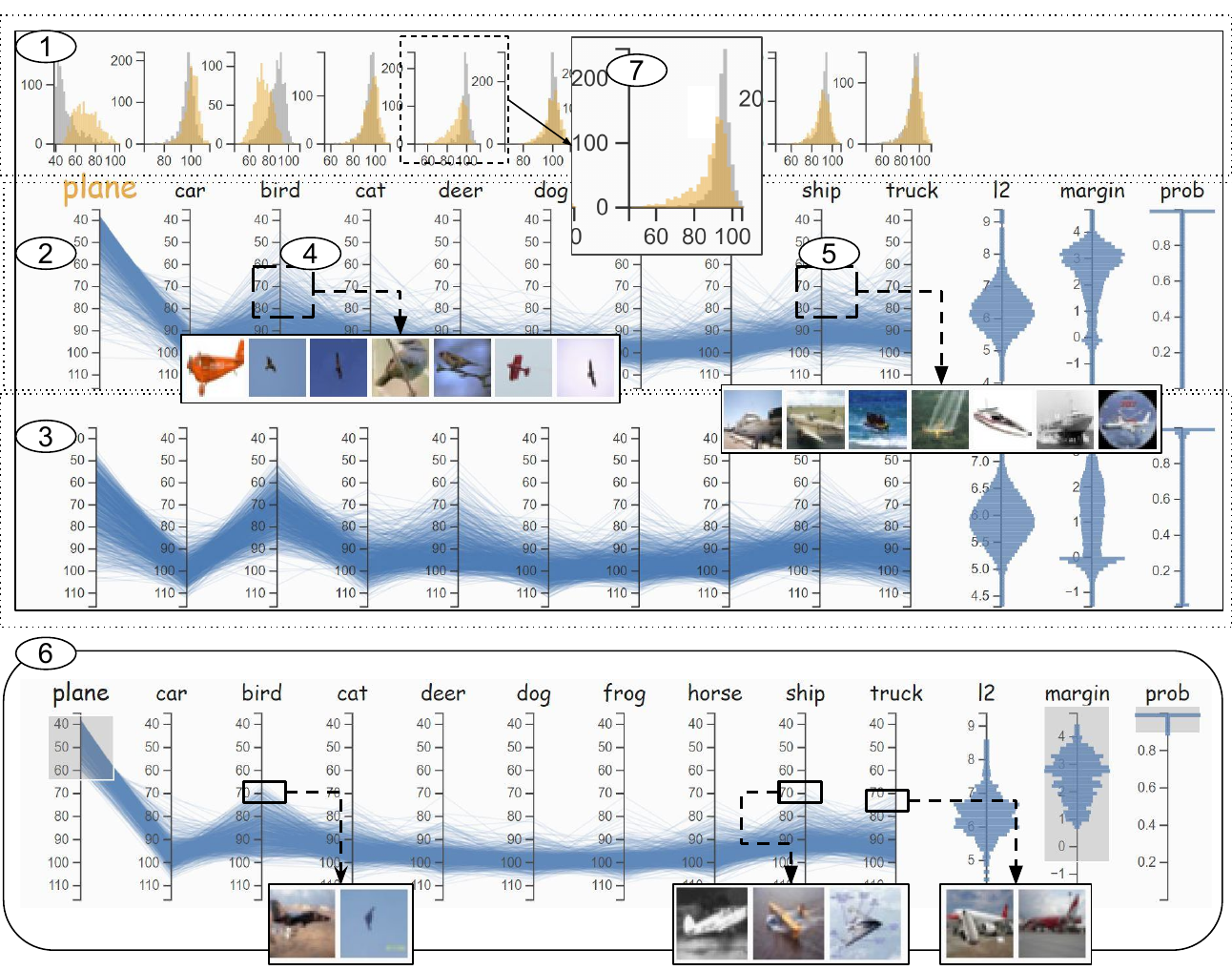}
   
    \vspace{-3mm}
    \caption{\major{VGG16 model's behavior over cifar10 plane samples, and the same data but corrupted with fog corruption. In the reference dataset \textcircled{\small{2}}, these samples are confused with bird and ship samples. In the fog-corrupted version \textcircled{\small{3}}, the confusion is strengthened. In \textcircled{\small{6}}, the visualization displays a set of samples that are predicted correctly with high probability but actually show certain ambiguities with the other labels.}}
\label{fig:global_geometry_comparison}
\vspace{-3mm}
\end{figure}

For example, Fig.~\ref{fig:global_geometry_comparison} \textcircled{\small{2}} displays samples belonging to the plane class.
Most of the samples have smaller angles with respect to the direction of the plane class in comparison to other classes.
However, these plane samples also have relatively small angles with the bird class (Fig.~\ref{fig:global_geometry_comparison} \textcircled{\small{4}}) and the ship class (Fig.~\ref{fig:global_geometry_comparison} \textcircled{\small{5}}), which can lead to incorrect classification.
% From these samples, we can see plane images and bird images have similar \emph{angle} metric, which may be fools the classifier.
By further examining these samples, we can see a similar shared background, which might be one contributing factor for this mistake.
\major{The other features (l2 norm, margin, and probability) are displayed as density plots that are separated from angle features because of the difference in semantic meaning and value scale.}

\major{Similarly with the local geometry plot, the global geometry plot also enables the geometric comparison among models and datasets (\textbf{R3}).
In Fig.~\ref{fig:global_geometry_comparison} \textcircled{\small{3}}, the visualization displays the same plane class but with samples that are corrupted with fog noise.
Before corruption, only a few plane samples had a small angle with the ship and the bird classes. 
However, with the presence of fog corruption, the number of samples that have small angles with both plane and bird increases. 
The density plot (Fig.~\ref{fig:global_geometry_comparison} \textcircled{\small{7}}) on the top of each class axis represents the distribution of the angle value. The x-axis means the number of samples and the y-axis is the angle of a specific class.
The gray color highlights the reference dataset (original), and the yellow color highlights the compared dataset (corrupted). In the visualization, the bird class samples' \emph{angle} metrics shift to the smaller values in the presence of corruption. Moreover, the corrupted dataset as a whole has a smaller \emph{margin} and \emph{length} compared to the clean datasets.
Such a result indicates that the uncertainty of the model's prediction between plane samples and bird samples increases.}
Our comparative interface allows us to extensively reason about similarities and differences among models pruned with various techniques. A detailed case study is presented in section 7.2.

A confusion matrix~\cite{hinterreiter2020confusionflow} aggregates many samples' prediction results to highlight the ambiguities between label predictions. 
However, if a sample is predicted correctly, the information from the confusion matrix will not help to tell whether the model may confuse this sample with other labels or not.
The output probability of a model is supposed to reveal the confusing information of a sample, but many studies have found that a model's output probability is often overcalibrated~\cite{guo2017calibration,xenopoulos2022calibrate}, and researchers should be cautious about how to read these probability values.
Compared with previous visual encoding, our design visualization and geometric metrics give more detailed information about a model's prediction.
In Fig.~\ref{fig:global_geometry_comparison} (\textcircled{\small{6}}), we demonstrate a set of plane samples that are predicted correctly with high probability, but have a related small angle with other class labels.
These samples are often fragile to data perturbation and pruning. The confusion analysis and output probability do not provide explanations for these vulnerabilities.

\subsection{Geometric Attribution View}

\begin{figure}[b]
\centering
    \includegraphics[width=\linewidth]{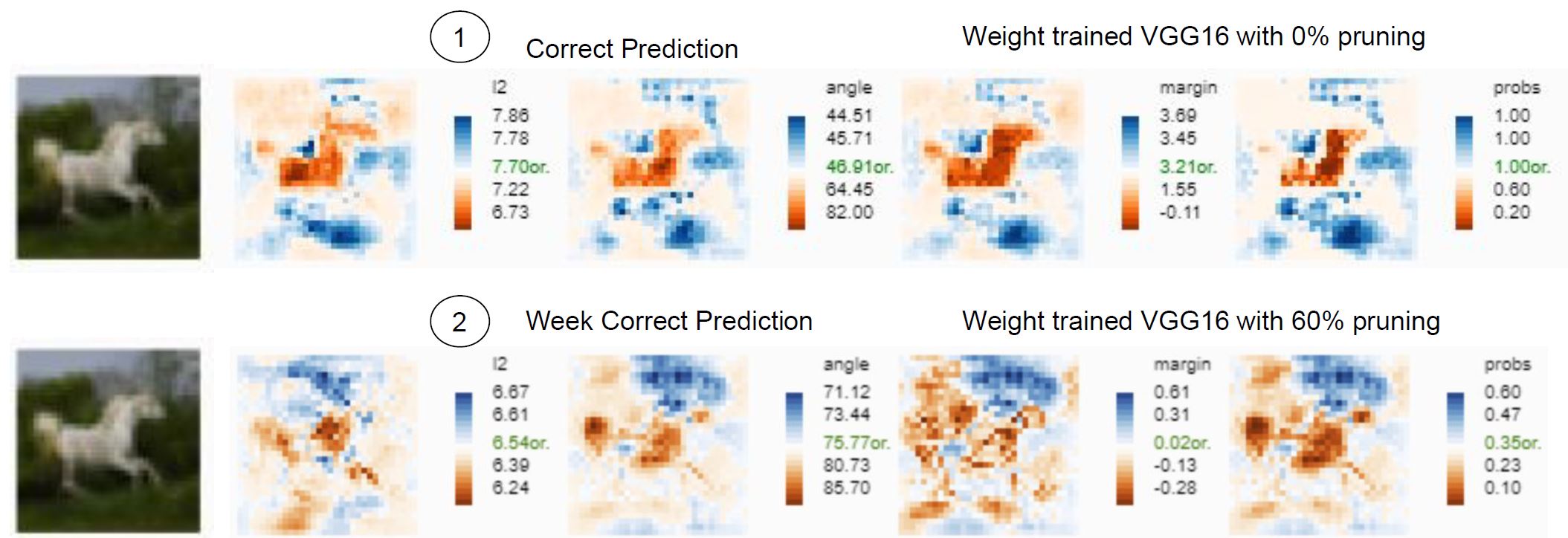}
   
    \vspace{-3mm}
    \caption{A geometric feature attribution view gives information about the sensitivity of input pixels of an image. From left to right are the sensitivity heat maps with respect to l2, angle, margin, and probability. From top to bottom are the results of VGG16 model with a different pruning ratios. Models with different pruning ratios will change the importance of an input image's pixel to the final decision.}
\label{fig:geometry_feature_attribution}
\vspace{-5mm}
\end{figure}

The geometric feature attribution view shows how each pixel of an input image contributes to different geometric features.
Our method is designed based on a perturbation technique that occludes a part of the input image and checks how much a model changes a sample's latent geometric features compared with the original image (\textbf{R5}).
The variation of these geometric feature values is visualized as a heatmap to highlight input pixels that can change.
During the design process, we have tried different gradient-based feature attribution methods~\cite{adadi2018peeking,simonyan2013deep} to compare the input features that are captured by different models.
However, we found that gradient-based feature attribution methods often highlight the same pixel even if the selected models have significantly different prediction performances, and this observation is consistent with previous research observations~\cite{adebayo2018sanity}. 
Alternatively, we use a perturbation-based approach to highlight the sensitive region for the model prediction.

In Fig.~\ref{fig:geometry_feature_attribution}, we demonstrate the geometric feature attribution of an image over three geometric metrics and probability from a well-trained VGG16 with different pruning ratios. 
For example, the heat map of the margin highlights the input pixels that are critical for prediction robustness.
The color map of each heat map is annotated with a metric value scale. The green text is the geometric value of an image without pixels occluding. 
Blue represents the increase of the geometric metric and red indicates a decrease of the value.
Removing the blue region will strengthen a certain geometric feature, and the red region will weaken a geometric feature.

The well-trained VGG16 in Fig.~\ref{fig:geometry_feature_attribution} (\textcircled{\small{1}}) predicts the image belonging to the horse category with high confidence.
Meanwhile, all four feature sensitivity views highlight the horse's body.
This result indicates that for this image, prediction confidence, margin, angle, and l2 norm features focus on similar input.
With 60\% weights pruned, the model still predicts the image as a horse correctly but with much lower confidence.
Compared with the original prediction, the new prediction's four feature sensitivity map highlights the inconsistent part of the image but with certain portions focused on the horse's body.
%Once 90\% of the weights pruned, the model make wrong predictions about this image and four heat maps highlight inconsistent part of the image. The margin feature sensitivity view highlight the part of the image completely irrelevant, 

\subsection{Coordination Between Views}
\major{In the system, five views coordinate with each other to address domain requirements. We summarize their relationship with each domain task in Table~\ref{tab:taskvscomponent}, and show how they interact with each other to expand the exploration ability of the visualization system in Fig.~\ref{fig:workflow}.}
\major{Users can interact between the geometric similarity view and evaluation table view to study and compare the similarity between evaluation benchmarks and pruning methods (Fig.~\ref{fig:workflow} (\textcircled{\small{1}})) and select interesting models for detailed examination and comparison (Fig.~\ref{fig:workflow} (\textcircled{\small{2}})).
During the pruning analysis, domain experts are also interested in the performance of different architectures and pruning techniques on a subset of samples or a specific class (Fig.~\ref{fig:workflow} (\textcircled{\small{3}})).
To achieve this requirement, during the downstream analysis, our system enables users to select a subset of samples, and the evaluation table can automatically reflect the result of current samples (\textbf{R3}).
The current visualization also enables a comparative analysis, which shows the performance comparison between currently selected samples and the entire test dataset. 
% In the visualization, the blue bar is the accuracy of model on cifar10 or cifar10c dataset.
The length of the green bar indicates that the performance increases on the selected subset of samples compared to the accuracy of the overall test dataset.
%For example, the test accuracy of a model is 90\% but the model's performance on the selected subset samples is 95\%, then the green bar indicate the performance improvement. 
The red bar shows the opposite concept revealing a decline in the performance. 
Such an operation enables users to examine more details of model behavior and helps them understand model performance under different levels of granularity. At the end, users can select a specific samples  for detailed feature attribution comparison (Fig.~\ref{fig:workflow} (\textcircled{\small{4}}))}

%\begin{figure*}[t]
%\centering
%    \includegraphics[width=\linewidth]{images/6_visualization_interaction}
   
%    \vspace{-3mm}
%    \caption{\major{The interaction between visual components of the system. Users can interact (\textcircled{\small{1}}) between evaluation table view and geometric similarity view to compare the similarity of pruning methods and evaluation benchmark. With selected interesting models and benchmarks, user can examine their detail geometric properties (\textcircled{\small{2}}) through local geometric view and global geometric view. In the meantime, user can select a subset of samples(\textcircled{\small{3}}) in local geometric view or global geometric view base on users' interests to examine their performance in different corruption benchmark and pruning methods. At the end, user can examine a detail sample's feature attribution (\textcircled{\small{4}}) to compare what features are captured by different models.}}
%\label{fig:visualization_interaction}
%%\vspace{-2mm}
%\end{figure*}

\begin{table}[t]
\vspace{-3mm}
\caption{Composition of visual components to address each domain task.}
\begin{center}
\vspace{-3mm}
\begin{tabular}{ |c|c|c|c|c|c|c|c| }
 \hline
 \multicolumn{2}{|c|}{Visual Components\textbackslash Tasks} & R1 & R2 & R3 & R4 & R5 \\ 
 \hline
\multicolumn{2}{|c|}{Evaluation Table View} &\checkmark& \checkmark & \checkmark  & &\\
  \hline
\multicolumn{2}{|c|}{Geometric Comparison View} &\checkmark& \checkmark &   & &\\
  \hline
\multicolumn{2}{|c|}{Local Geometry View} & &  & \checkmark & \checkmark &\\
  \hline
\multicolumn{2}{|c|}{Global Geometry View} &  &  &   \checkmark & \checkmark&\\
 \hline
\multicolumn{2}{|c|}{Geometric Attribution View} &  &  &  &  & \checkmark\\
 \hline
\end{tabular}
\label{tab:taskvscomponent}
\end{center}
\vspace{-3mm}
\end{table}

\section{Use Cases}

\begin{figure*}[t]
\centering
    \includegraphics[width=\linewidth]{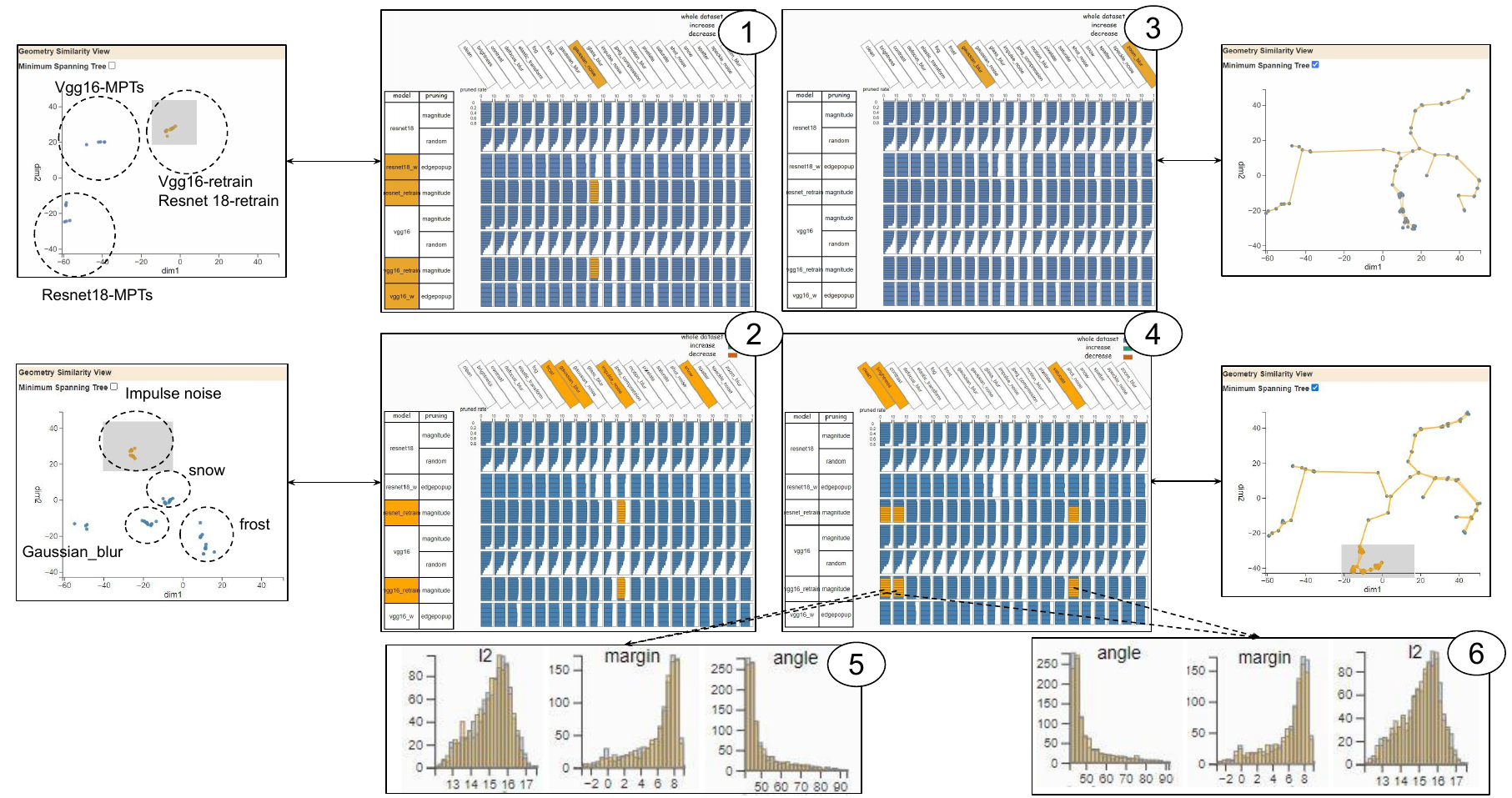}

    \vspace{-3mm}
    \caption{\major{In the visualization, \textcircled{\small{1}} reveal that MPTs pruning approach and magnitude pruning approach end up with different geometric feature representations over Gaussian noise corruption.The finding of \textcircled{\small{2}} reveals that pruning and retraining with Vgg16 and Resnet18 models have highly comparable geometric feature representations. \textcircled{\small{3}} and \textcircled{\small{4}} together emphasize the existence of redundancy in common corruption evaluation benchmarks. \textcircled{\small{3}} shows redundancy between the Gaussian blur and the zoom blur benchmarks. \textcircled{\small{4}} highlights the similarity among clean, \emph{brightness} and \emph{saturate} benchmark. In \textcircled{\small{5}}, \textcircled{\small{6}}, we select two pairs of evaluation benchmarks on the vgg16 retrain model for comparison. Their geometric features' distribution are similar and overlap with each other.}}
\label{fig:similarity_of_evaluation_benchmark}
\vspace{-4mm}
\end{figure*}

\major{In this section, we demonstrate the usability of our system by demonstrating how it can provide valuable insights and impactful answers regarding behaviors of various pruning methods, the efficiency of evaluation benchmark, and the robustness of the pruned models to common data corruptions.}

\subsection{Comparing Network Pruning Approaches and Identifying Similar Evaluation Benchmarks}

\major{Exploring and testing the performance of network pruning approaches' over multiple corruption evaluation benchmarks are often the initial tasks for domain experts to understand their differences.
This process involves questions such as whether different pruning approaches will yield similar or different behaviors. 
The answer to this question helps researchers understand the pros and cons of different pruning methods.
Domain experts tell us that they often use similarity metric centered kernel alignment (CKA)~\cite{kornblith2019similarity} to compare two pruned models' feature representations one at a time.
However, as the number of feature representations for comparison increases, this approach becomes increasingly intractable. 
Therefore, there is a growing need for a scalable methodology that can efficiently assess the similarity among multiple models simultaneously.}

\major{In Fig.~\ref{fig:similarity_of_evaluation_benchmark}, we illustrate how domain experts employ a combination of the geometric similarity view and evaluation table view to address the above concerns. 
In Fig.~\ref{fig:similarity_of_evaluation_benchmark} \textcircled{\small{1}}, we select the data from the Gaussian noise corruption benchmark to generate the feature representations of different models.
Specifically, we apply magnitude and MPTs pruning to VGG16 and Resnet18 network architectures for comparative analysis.
From the visualization, we can tell that magnitude pruning, when applied with retraining, on both VGG16 and Resnet18 results in feature representations that have a similar geometry as their dense counterparts.
Conversely, MPT pruning over these two network architectures makes significant geometric changes.
One intriguing observation of \textcircled{\small{1}} is that retrained VGG16 and Resnet18 have comparable feature representation over the Gaussian noise corruption benchmark.
However, MPT pruning applied to VGG16 and Resnet18 led to significantly different feature representations.
To further validate the consistency of similarity between retrain VGG16 and Resnet18, we narrow our focus to these two models and evaluate them with additional corruptions such as snow, Gaussian blur, frost, and impulse noise.
In Fig.~\ref{fig:similarity_of_evaluation_benchmark} \textcircled{\small{2}},
the resulting visualization reveals that the similarities between these two models are consistent across multiple evaluation benchmarks.}

\major{During the comparison and evaluation, understanding the coverage of the evaluation benchmark is important to assess the quality of the analysis.
In the evaluation table view, there are 20 distinct evaluation benchmarks, and each of them is created by different corruption algorithms~\cite{hendrycks2019benchmarking}. 
During the design of these corruption algorithms, whether these evaluations will lead to similar model behavior has not been thoroughly explored.
Corruption operations that result in a similar model behavior may be redundant and unnecessary.
In Fig.~\ref{fig:similarity_of_evaluation_benchmark} \textcircled{\small{3}},\textcircled{\small{4}}, we show how such similarity can be revealed by our system.
\textcircled{\small{3}} highlights two corruption benchmarks, Gaussian blur, and zoom blur, in the evaluation table view and geometric similarity view.
Notably, the minimum spanning tree constructed by these models evaluated with the selected benchmarks exhibits a significant overlap with each other. 
Similarly, in \textcircled{\small{4}}, the minimum spanning trees of the models that are evaluated with the corruption evaluation over brightness, saturation, and the clean dataset also overlap substantially.
We can further use the local geometric view to compare their local geometric feature variation.
In \textcircled{\small{5}} compares the geometric features distribution of retrain vgg16 over clean and the brightness corruption data, and \textcircled{\small{6}} compares the clean and the saturate corruption data.
In both cases, the difference between their geometric distribution is subtle.
In \textcircled{\small{5}} and \textcircled{\small{6}}, each histogram has two color distributions: orange and steel blue. Because these two histograms are highly overlapped,  only one is color histogram displayed at the end.
These results suggest that for the current image dataset and tasks, models subjected to the currently available pruning methods have similar reactions to the evaluation benchmark.
Consequently, there is an opportunity for optimization, such as removing similar benchmarks, to streamline the analysis process, reduce computation complexity, and minimize cost.}

\begin{figure*}[t]
\centering
    \includegraphics[width=\linewidth]{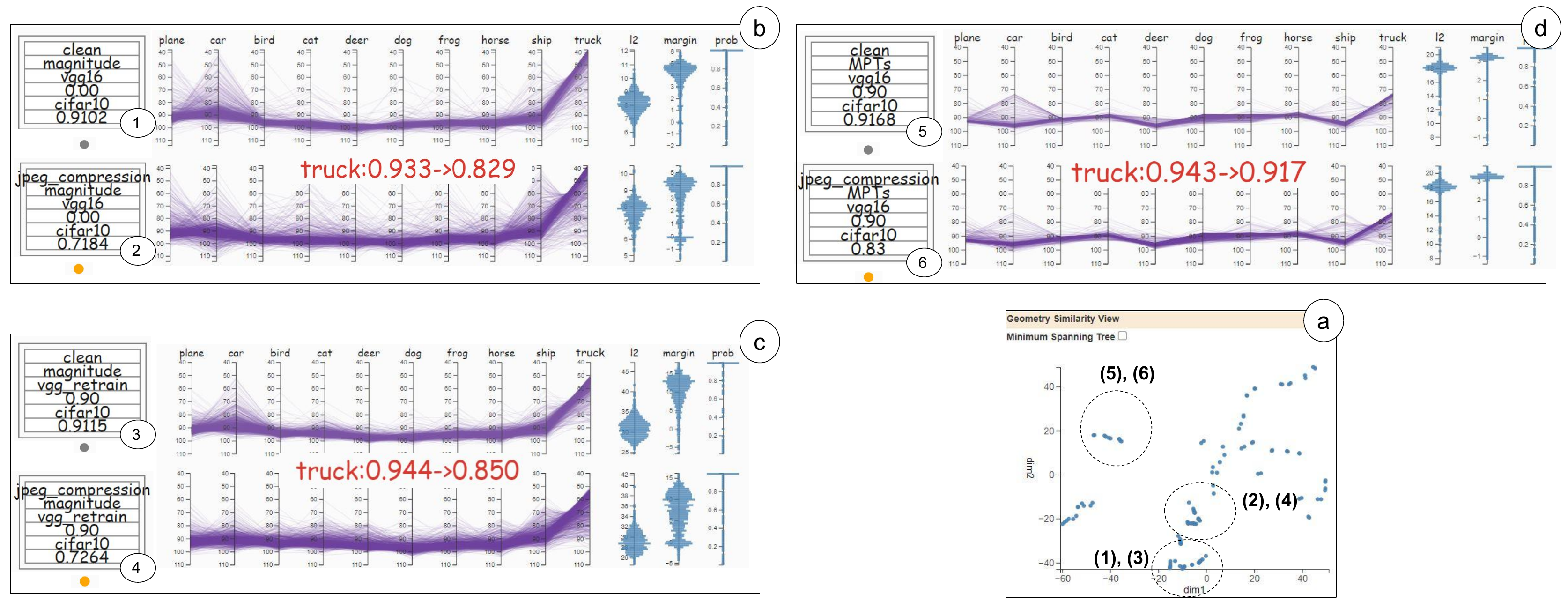}
    \vspace{-6mm}
    \caption{\major{The geometric similarity view \textcircled{\small{a}} highlights the position of related models. The visualization result reveals geometric disparities in their feature representation.
    The panel \textcircled{\small{c}} presents a comparison between truck samples of cifar10 on a 90\% weight-pruned retrained VGG16, and the same samples but corrupted with the JPEG compression. 
    The panel \textcircled{\small{b}} showcases a dense weight well-trained VGG16 and its representation of clean and corrupted data.
    The same comparison \textcircled{\small{d}} is performed on the VGG16, which is generated by the MPT method. 
    The JPEG corruption induces pronouncedly more angular variations with multiple class directions and minimum distant shift in the weight-trained VGG16 and retrained VGG16 than the VGG16 generated by the MPTs.
    This observation provides valuable insights into why the models generated by the MPTs' method tend to be more robustness to data corruption compared to the regular weight-trained model and weight-pruned retrained model.}}
\label{fig:use_case_7_2}
\vspace{-2mm}
\end{figure*}

\subsection{What Is the Representation Difference Between the Robust MPTs Pruned Models and Others (Dense and Retrained)?}

\major{The use case in section 7.1 gives hints about the reaction of different pruned models over multiple evaluation benchmarks.
However, it does not give details about the difference between retrained models and MPTs pruned models.
The MPTs method has been proven~\cite{diffenderfer2021multiprize} to produce compact models, which not only have high prediction accuracy and small model size but also models that are significantly more robust to various data corruptions than regular weight-trained models\footnote[1]{\href{https://robustbench.github.io/div\_cifar10\_corruptions\_heading} \url{https://robustbench.github.io/div\_cifar10\_corruptions\_heading}}.
However, it is still unclear why and how the MPT approach achieves such a performance gain. 
Here, we show that our visualization tool can help develop hypotheses about answering this question by comparing the latent spaces' geometric structures of the traditional weight-trained VGG16, the retrained VGG16 model with a certain pruning ratio, and the VGG16 model generated by MPT.}

\major{From Fig.~\ref{fig:use_case_7_2} \textcircled{\small{a}}, the visualization result reveals the significant geometric disparity of the model's feature representation.
The related models are numbered for reference.
To examine this in more geometric detail, we use the global geometric view to understand the difference between models.}

In Fig.~\ref{fig:use_case_7_2}, the panel (\textcircled{\small{b}}) shows the geometric difference of samples from the truck class with and without the JPGE corruption in the original weight unpruned trained VGG16 model.
The model is trained on the cifar10 dataset with 200 epochs, and the final accuracy on clean test data is 91\%. 
With the same test dataset but corrupted with the JPEG compression, the accuracy of the model drops to 72\%. 
The models are able to distinguish truck images from samples of other classes even though some samples may be slightly confused with car samples (\textcircled{\small{1}}). 
Once the dataset is corrupted, the model displays confusion with multiple categories such as plane/ship (\textcircled{\small{2}}), and the models' global geometric features on corrupted data are not as coherent as the clean dataset.
The samples belong to the truck class, and the prediction accuracy dropped from 93.3\% to 82.9\%.

%The local geometric visualization on the right shows a summary of the geometric feature distribution shift of the truck samples. The overall accuracy drops from 93.3\% to 82.9\%.
%The \emph{margin} of samples to the decision boundary, \emph{angle}, and \emph{length} of the samples have a noticeable shift. This pattern is expected as the model has not seen the corrupted data during training. 

The panel (\textcircled{\small{d}}) illustrates the same comparison, but using the models generated by the MPTs method, i.e., starting with an untrained VGG16 model and then 90\% of the weights are pruned, resulting in a model with 91\% accuracy. 
The performance of the model declines to 83\% when tested on the corruption dataset, which is significantly better than the performance of the dense weight trained model (71.84\%). 
For samples belonging to the truck class, the performance of the model only slightly drops from 94.3\% to 91.7\%.
The \emph{angle} distribution of samples with each class direction displays a distinct pattern, i.e., the dense VGG16 has a much larger variance but smaller mean \emph{angle}, whereas the MPT model has much a smaller variance but larger mean.
For MPTs, the global geometric visualization comparison \textcircled{\small{5}},\textcircled{\small{6}} shows that the difference between corrupted and clean data is minor.% and the relative local geometric visualization reveals that these three geometric distribution have much less distribution shift compared with \textcircled{\small{1}},\textcircled{\small{2}}. 

A simple potential explanation of such an observation is that the pruned model contains fewer parameters, which may lead to a latent space that contains much less information, and the model's prediction should be less sensitive to the input noise, which leads to a more robust model.
To verify this hypothesis, we add an additional comparison (\textcircled{\small{c}}), in which the model is pruned with 90\% weight but retrained 50 epochs to gain the original prediction accuracy around 91.14\%.
For the truck class, the prediction accuracy increases to 94.4\%, and the accuracy of the corrupted data increases from 82.9\% to 85.0\%. 
The new comparison gives positive feedback about the hypothesis that decreasing the number of parameters in a model can improve a model's resiliency to data corruption.

However, the model generated by the MPTs model's prediction accuracy on corruption is still significantly better than the retrained model.
This behavior indicates that besides the number of parameters in the model-generated MPTs, the unique geometric latent space of the MPTs model can play a significant role.

After presenting this use case to the domain experts, we collected and summarized their feedback about how this information is helpful to their research as follows: 
"Such an observation motivates a hypothesis that the representation geometry of MPTs is less sensitive to different corruption types than that of the regular weight-trained or retrained models.
These sensitive properties may related to the thin standard deviation of geometric features and angle scales of different models' feature representations.
Meanwhile, this finding has potentially significant implications for robust machine learning as it suggests that to design a robust neural network, one should not only optimize for the training accuracy but also incorporate additional geometric constraints during training".

\begin{figure*}[t]
\centering
    \includegraphics[width=\linewidth]{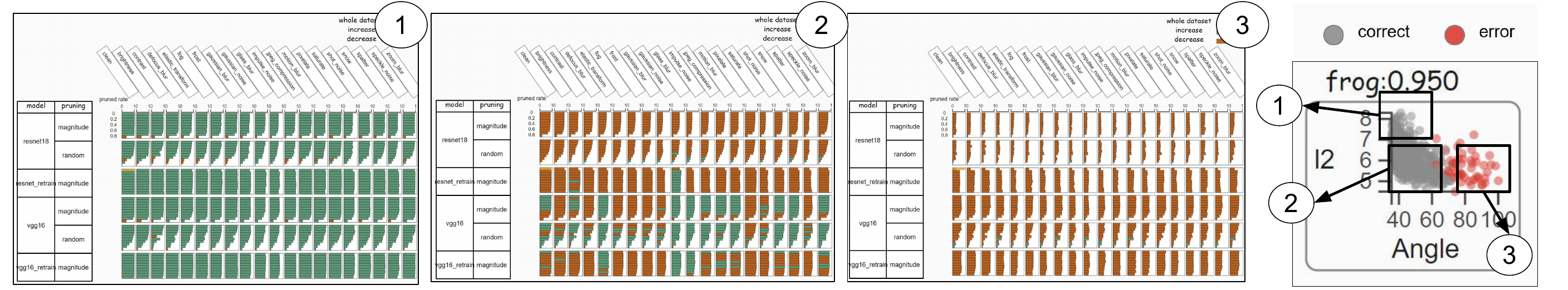}
    \vspace{-7mm}
    \caption{An overview of the evaluation comparison across multiple architectures and pruning techniques of frog samples. The sample with large \emph{length} and small \emph{angle} are less affected by the pruning, different model architecture, and data corruption.
    On the other hand, the samples with smaller \emph{length} and large \emph{angle} are fragile over different pruning approaches and models.}
\label{fig:use_case_7_3}
\vspace{-1mm}
\end{figure*}

\subsection{How Do Samples' Geometry Impacted by Model Pruning, Data Corruptions, and Model Retrain?}
\major{Model evaluation often aggregates a model's prediction over all test samples, but often misses critical details about the model's behavior in a specific category or a subset of samples. 
The previous network pruning literature~\cite{hooker2019compressed} has assessed the impact of network pruning over samples by proposing a metric called PIE (pruning-identified exemplars) to identify vulnerable samples.
The PIE value is calculated by the disagreement between compressed and uncompressed models that require a large amount of computation resources.
At the same time, this metric does not reveal what properties cause a sample's vulnerability. 
Here, we demonstrate how we can use our visualization system to quickly highlight these samples and provide depth analysis such as how pruning and retraining affect a sample and what features are captured or forgotten by the model.}
%Meanwhile, understanding how pruning impact model behavior is important to  how retrain affect a single sample and how its feature is changed by pruning operation.

\major{In Fig.~\ref{fig:use_case_7_3}, we demonstrate an interactive exploration case that compares three subsets of samples' observations with different geometric properties.
We select these samples from the frog category of the cifar10 dataset and evaluate their performance on pruned models over available evaluation benchmarks. 
Fig.~\ref{fig:use_case_7_3}\textcircled{\small{1}} displays the evaluation result of samples that are selected from the frog category with large \emph{length} and small \emph{angle}.
In the evaluation table view, we can tell that these samples perform better than the whole dataset with high prediction accuracy with different data corruption benchmarks.
Meanwhile, they are less affected by the different pruning techniques.}

\major{Compared with Fig.~\ref{fig:use_case_7_3}\textcircled{\small{1}}, Fig.~\ref{fig:use_case_7_3}\textcircled{\small{2}} is the evaluation result from samples that have a relatively larger angle and smaller l2 norms.
The overall performance of these samples declines compared with samples from  Fig.~\ref{fig:use_case_7_3}\textcircled{\small{1}}. 
These samples demonstrate less resiliency to different pruning and corruption operations.
The last case is Fig.~\ref{fig:use_case_7_3}\textcircled{\small{3}}, which shows samples with small l2 and a large angle. 
All these samples have poor performance in different evaluations and different model architectures.
Meanwhile, the Resnet18 model performs worse than the VGG16 model over these samples under different pruning techniques and retraining models.}

\major{The above observation reveals that large l2 and small angle values often show resiliency to the different pruning operations across multiple network models and different data corruption evaluations.
On the other hand, samples with small \emph{length} and large \emph{angle} values display fragile behavior, and these samples are affected dramatically by pruning and data corruption.
To further verify such observations, we follow domain experts' suggestion to perform a quantitative evaluation to measure the relationship between geometric properties and data corruption.}

\major{Our experiment measures two relationships: how does a sample's geometric properties correlate with different data corruptions, and how does a sample's geometric properties correlate with standard model pruning?
In Table 1 and Table 2, we demonstrate our evaluation result over the Imagenet dataset.
The result finds that the angle and margin metric are strongly correlated with the samples' resiliency to data corruption and model pruning.
However, the l2 norm does not show a consistent correlation relationship between pruning and data corruption. The support material includes details of how we performed the experiment and more dataset evaluation results.
This result shows that geometric properties can be used as a metric to highlight vulnerable and resilient samples without comparing many models, and it is a significant advantage compared with the previous approach~\cite{hooker2019compressed}.}

\begin{table}[t]
\centering
\begin{tabular}{cccccc}
 \hline
CNN Architecture  & rc-angle   & rc-l2 &  rc-margin. \\  \hline
 VGG16     & -0.636 & 0.516 & 0.685 \\  
 resnet18      &  -0.715 & 0.192 & 0.696\\
 resnet50      & -0.7172 & 0.053 & 0.6677\\
 resnet152      & -0.7158 & -0.035 & 0.7228\\  
 densenet121      & -0.726 &  -0.015 & 0.6732\\  \hline
\end{tabular}
\caption{The result of the \textbf{Imagenet-C} validation dataset. This table shows the Pearson correlation coefficient between different geometric features and models' robustness. Angle and margin show a significant correlation with robustness. The correlation between l2 and robustness is moderate or subtle.}
\vspace{-2mm}
\end{table}

\begin{table}[t]
\centering
\begin{tabular}{cccccc}
 \hline
CNN Architecture  & rc-angle   & rc-l2 &  rc-margin. \\  \hline
 VGG16     &  -0.6739 & 0.3924 & 0.6534 \\  
 VGG19    &  -0.6667 & 0.3984 & 0.6514 \\  
 resnet18      & -0.6795 & 0.1737 & 0.7282\\
 resnet50      & -0.6933 & 0.083 & 0.7058\\ 
 Densenet121      & -0.6816 &  0.0252 & 0.7154\\\hline
\end{tabular}
\caption{The result of the \textbf{Imagenet} validation dataset. This table shows the Pearson correlation coefficient between different geometric features and model magnitude pruning. Angle and margin show a significant correlation with pruning vulnerability. The correlation between l2 and pruning vulnerability is subtle.}
\vspace{-2mm}
\end{table}

\begin{figure}[t]
\centering
    \includegraphics[width=\linewidth]{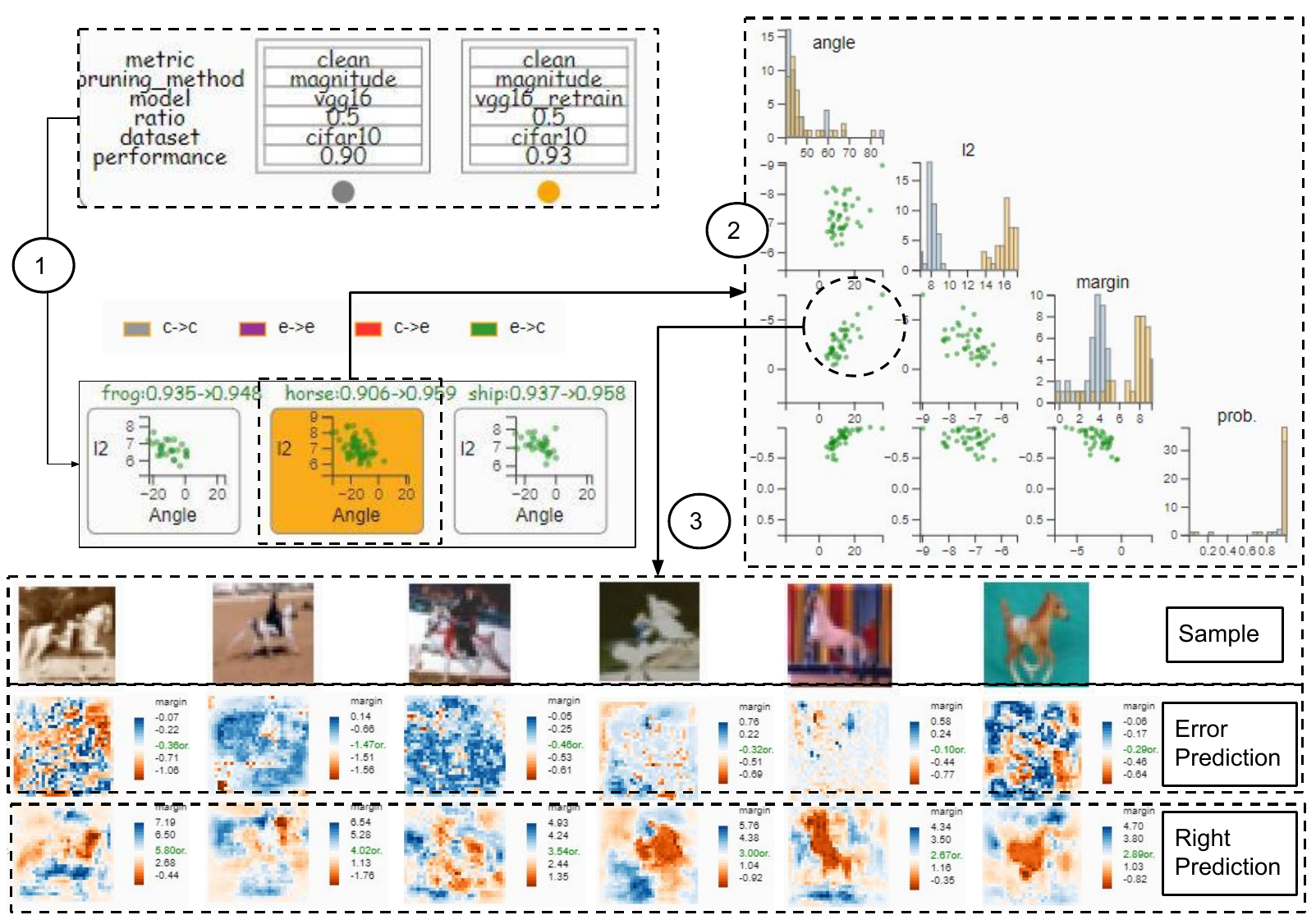}
    \vspace{-4mm}
    \caption{Model pruning and retraining can improve prediction performance. Comparing the retrained and original model (\textcircled{\small{1}}), the samples that are predicted wrong in the original model but predicted right in the retrained model have captured better features with respect to margin. (\textcircled{\small{3}}).}
\label{fig:use_case_7_2_2}
\vspace{-1mm}
\end{figure}

\major{Model pruning and retraining not only recover the performance of the pruned model but also improve it, especially for certain data samples.
We can check the feature attribution heat map with respect to the margin in Fig.~\ref{fig:use_case_7_2_2}. 
We compare the geometric feature attribution heat map of the pretrained model and retrained model with $50\%$ weight removed.
We select Fig.~\ref{fig:use_case_7_2_2}(\textcircled{\small{1}}) samples (green color) that are predicted incorrectly in the pretrained model but correctly in the retrained model.
The geometric feature attribution visualization in Fig.~\ref{fig:use_case_7_2_2}(\textcircled{\small{3}}) shows that the retrained model captures better features than the pretrained model.}
\section{Discussion and Conclusion}

%\textcolor{red}{Add a discussion about the similarity between feature representation across layer. lead to the potential next step of visualization compare previous layer's feature representation}

In this work, we introduced three geometrically inspired metrics in the neural network latent space for a comparative study of how widely adopted (and state-of-the-art) model pruning approaches impact neural network models' internal representation and performance.
% In particular, we examined the robustness of pruned models, and provide explanation on why certain pruning methods produce surprisingly robust model while other reduce the model's robustness.
% We design a novel visualization to understand and comparing the latent space of the network network model pruned with different pruning strategies.
The proposed visualization system is able to highlight the key differences between pruning techniques that are unknown to ML researchers. 
The visualization also provided valuable insights for explaining model robustness from a geometric perspective and answered why certain pruning methods produce surprisingly robust models whereas others reduce model robustness.

% Classical visualization technique T-SNE~\cite{van2008visualizing} for model latent space behavior study highlights the local nearest neighbors information of feature vectors during model prediction. % However, these information along is not sufficient to explain the model behavior such as why it makes mistake, why the model is more confident on some samples than the other, and why the model can be more  robust than the other. In comparing, our visualization incorporates with three geometric features and gives a global overview of the model's reaction to the samples. It visually displays the confusions between samples and give an geometric explanation of why these confusion exists.
%We incorporate these feature into an interactive visualization that enable machine learning engineers to compared the pros and cons of different pruning techniques and model architectures.

One potential limitation of the current system is the scalability of the parallel coordinate view.
As for more complex datasets, such as Imagenet (1000 classes), visualizing all \emph{class directions} at the same is not practical.
However, practically speaking, it is important to note that even with a more scalable visual encoding, there is a limit to how many classes a user can meaningfully examine at the same time. %due to the cognitive hurdle of humans.
Accordingly, a pres-election or ranking operation can greatly mitigate this challenge. 
In the future, we plan to develop an easy-to-use interface to help users select a subset of the interesting classes with different criteria, e.g., classes mostly influenced by network pruning or data perturbation.
Another potential improvement of the current system for future work comes from additional introspection ability, i.e., can we combine other model explanation tools such as saliency map \cite{selvaraju2017grad} or concept-based explanation \cite{kim2018interpretability} with the proposed geometry metric, to better articulate the exact semantics changes induced by pruning?

\newpage
\section{Supporting Materials}
In this section, we describe the quantitative evaluation of three geometric metrics.
We measure the correlation between these geometric metrics and data corruption.
Further, we also measure the correlation between the impact of magnitude pruning and these geometric metrics.
We find that both angle and margin are highly correlated with corruption operation and model pruning.
Finally, we demonstrate that the loss functions have a strong impact on the geometric shape of a neural network's feature representation.
We perform our experiment on MNIST, Cifar10, Cifar100, and ImageNet datasets. 
For the MNIST dataset, we train LeNet5 with 50 epochs.
The model trained with the cifar10 and cifar100 datasets is trained with 200 epochs.
For the ImageNet dataset, we use the pretrained model from the PyTorch library to perform the experiments

\subsection{Correlation Between Geometric Metrics and Corruption Robustness }
To attribute the robustness to three geometric features, we perform an experiment to measure the correlation between corruption robustness and geometric properties. Note that the robustness is defined as the average accuracy on corrupted data (i.e., applying $c=19$ corruption types and $s= 5$ severity levels to each clean test image $x_i$ and transforming it to a set of corrupted images $x_i^{c,s}$):

\begin{align}
    robustness_i = \sum_{c=1}^{19} \sum_{s=1}^5 \mathcal{I}_{\{y_i = f(x_i^{c,s})\}},
\end{align}
where $\mathcal{I}$ is an indicator function taking value 1 if the prediction is correct on the corrupted image and 0 otherwise. Furthermore, $robustness_i$ denotes per sample robustness quantifying the level of invariance to corruptions. 

The result of our experiment is summarized in Tables 1, 2, 3, and 4, in which rc-angle, rc-l2, and rc-margin denote the Pearson correlation coefficient between the robustness and the angle, length, and margin, respectively.
Overall, the angle and margin to the decision boundary show a strong correlation with the corruption robustness.
However, the length is moderately correlated.
The correlation between length and robustness is not consistent across different architectures.
ResNet Architecture has a much smaller l2 correlation compared with regular convolution neural networks.

\begin{table}[b]
\centering
\begin{tabular}{cccccc}
 \hline
CNN Architecture  & rc-angle   & rc-l2 &  rc-margin. \\  \hline  
 Lenet 5      & -0.6902 & 0.3516 & 0.6359\\\hline
\end{tabular}
\caption{The result of the \textbf{MNIST-C} dataset. This table shows the Pearson correlation coefficient between different geometric features and models' robustness. Angle and margin show a significant correlation with robustness. The correlation between l2 and robustness is moderate or weak.}
\vspace{-2mm}
\end{table}  

\begin{table}[b]
\centering
\begin{tabular}{cccccc}
 \hline
CNN Architecture  & rc-angle   & rc-l2 &  rc-margin. \\  \hline
 Resnet18     & -0.8148 & 0.3401 & 0.7794 \\  
 VGG16      & -0.8107 &0.5783 & 0.8234\\  
 Alexnet     & -0.8572  & 0.5557 & 0.7681\\  \hline
\end{tabular}
\caption{The result of the \textbf{cifar10-C} dataset. This table shows the Pearson correlation coefficient between different geometric features and models' robustness. Angle and margin show a significant correlation with robustness. The correlation between l2 and robustness is moderate or weak.}
\vspace{-2mm}
\end{table}

\begin{table}[t]
\centering
\begin{tabular}{cccccc}
 \hline
CNN Architecture  & rc-angle   & rc-l2 &  rc-margin. \\  \hline
 VGG16     & -0.7665 & -0.0618 & 0.6891 \\  
 resnet18      & -0.7673 & 0.3334 & 0.7316\\
 Googlenet      & -0.7686 & 0.3815 & 0.7228\\\hline
\end{tabular}
\caption{The result of the \textbf{cifar100-C} dataset. This table shows the Pearson correlation coefficient between different geometric features and models' robustness. Angle and margin show a significant correlation with robustness. The correlation between l2 and robustness is moderate or weak.}
\vspace{-2mm}
\end{table}

%\BK{There is no need to write numbers in table in text. Delete this para: The first row displays the result of Resnet18.
%The correlation factor between angle and robustness is -0.814, and the correlation factor between angle and minimum distance to decision boundary is 0.7794. 
%Similar trend can also be seen in second row, which shows the result of VGG16, and the correlation factor is -0.8107 and 0.8234. 
%The third row shows the result of Alexnet, and the correlation factor is -0.8572 and 0.7681.  
%The correlation factor of l2 with robustness over Resnet18, VGG16 and Alexnet is 0.3401, 0.5783, and 0.5557.}

%We use the cifar10c dataset to measure the correlation between the robustness and the three global metrics. 
%To measure how the common data corruption will affect the global features.
%We perform our experiment with VGG16 which trained on cifar10 dataset, and use the cifar10-c as a corruption dataset to understand and compare different data corruption methods impact. Fig.~\ref{fig:corruption_summary} show a summary how data corruption will affect the examples' feature embedding. These corruption will increase the angle of sample from the target semantic direction, decrease the l2 norm of the feature embedding and reduce the minimum distance to the decision boundary.
%The l2 norm observation is interesting, as the data corruption only make the l2 norm smaller not bigger indicate that ambiguity region is more close to the original point.
\subsection{Correlation Between Geometric Metric and Pruning Vulnerability}
To measure the correlation between geometric metrics and pruning vulnerability (PV),  we use magnitude pruning with a different pruning ratio on the same model to understand the impact of magnitude pruning on different samples.
The operation is performed 10 times with the pruning ratio from $0\%$ to $90\%$ (e.g.,10\%, 20\%,...), which includes 10 ratios.
We collect each sample's correctness in each pruned model and aggregate the result as a $PV$ value. 
A large $PV$ value indicates that the samples' features are resilient to model pruning.
We measure the correlation between the $PV$ and the three geometric metrics.
\begin{align}
    PV_i = \sum_{s=1}^{10} \mathcal{I}_{\{y_i = f(x_i^{c,s})\}},
\end{align}

\begin{table}[t]
\centering
\begin{tabular}{cccccc}
 \hline
CNN Architecture  & rc-angle   & rc-l2 &  rc-margin. \\  \hline  
 Lenet 5      & -0.399 & 0.238 & 0.486\\\hline
\end{tabular}
\caption{The result of the \textbf{MNIST} dataset. This table shows the Pearson correlation coefficient between different geometric features and model magnitude pruning.}
\vspace{-2mm}
\end{table}

\begin{table}[b]
\centering
\begin{tabular}{cccccc}
 \hline
CNN Architecture  & rc-angle   & rc-l2 &  rc-margin. \\  \hline
 Resnet18     & -0.787 & 0.3067 & 0.7264 \\  
 VGG16      & -0.838 & 0.395 & 0.764\\  
 Alexnet     & -0.739  & 0.442 & 0.642\\  \hline
\end{tabular}
\caption{The result of the \textbf{cifar10} dataset. This table shows the Pearson correlation coefficient between different geometric features and model magnitude pruning. Angle and margin show a significant correlation with robustness. The correlation between l2 and robustness is moderate or weak.}
\vspace{-2mm}
\end{table}

\begin{table}[t]
\centering
\begin{tabular}{cccccc}
 \hline
CNN Architecture  & rc-angle   & rc-l2 &  rc-margin. \\  \hline
 Resnet18     &  -0.814 & 0.333 & 0.822 \\  
 VGG16      & -0.8731 & 0.002 & 0.8797\\  
 Googlenet     & -0.838  & 0.3687 & 0.836\\  \hline
\end{tabular}
\caption{The result of the \textbf{cifar100} dataset. This table shows the Pearson correlation coefficient between different geometric features and model magnitude pruning. Angle and margin show a significant correlation with robustness. The correlation between l2 and robustness is moderate or weak.}
\vspace{-2mm}
\end{table}

We perform our experiment on the MNIST, Cifar10, Cifar100, and Imagenet datasets. 
The result of our experiment is summarized in Tables 6, 7, and 8, in which rc-angle, rc-l2, and rc-margin denote the Pearson correlation coefficient between the pruning vulnerability and the angle, length, and margin, respectively.
Overall, the angle and margin to the decision boundary show a strong correlation with the pruning vulnerability.
Samples with small and large margins are more resilient to magnitude pruning than the other samples.
However, the length is a weak correlation.

\subsection{The Effect of the Curse of Dimensionality}\label{sec:curse_of_dimension}
\begin{figure}[htp]
\centering
    \includegraphics[width=\linewidth]{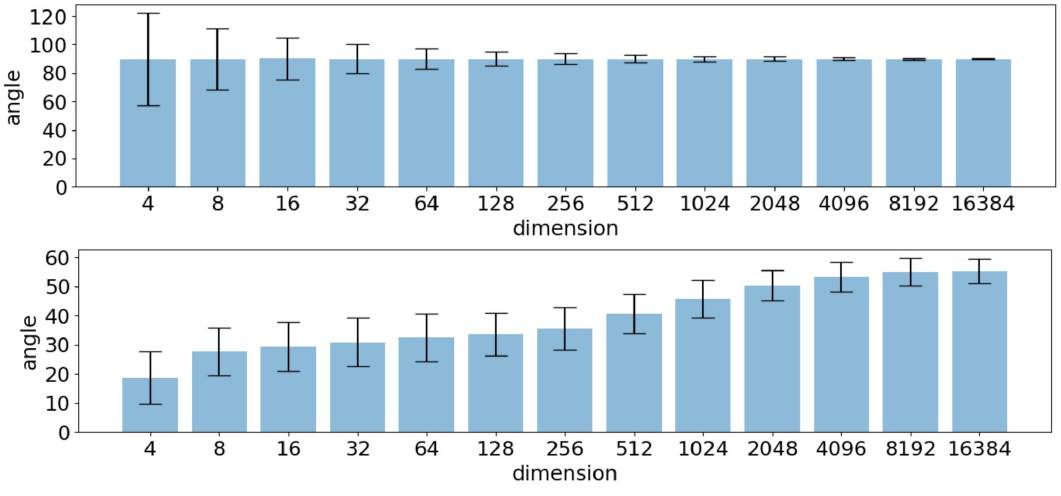}
    \vspace{-3mm}
    \caption{Top plot shows that with increasing dimensions, the angle between two uniformly sampled random vectors will converge to 90 degrees with smaller standard deviations. The bottom plot shows that increasing the number of neurons in the last feature layer will lead to a large angle between samples and their class direction. }
\label{fig:curse_of_dimensionality_angle}
\vspace{-2mm}
\end{figure}

High-dimensional geometry is affected by the curse of dimensionality~\cite{verleysen2005curse} such that the intuitions humans have on two- or three-dimensional spaces may not apply to high-dimensional space.
For example, the angle between two random vectors in high dimensional space will display counter-intuitive behavior.
In Fig.~\ref{fig:curse_of_dimensionality_angle}, the top plot is the result of an experiment that is performed on vectors generated by uniform random sampling. 
With the increasing dimension, the mean angle between 10000 uniformly random sample vectors will converge to 90 (orthogonal) and their standard deviation will become small. 
The orthogonality between two vectors indicates that they do not contain any information about each other.

The phenomenon also affects the high-dimensional latent space of well-trained neural network models. For untrained LeNet\_5 models, the samples' mean angle and standard deviation with the class direction are $88.8 \pm 6.8$, which indicates an uninformative feature vector.
The bottom plot of Fig.~\ref{fig:curse_of_dimensionality_angle} displays the average angle and standard deviation of the MNIST dataset with their class directions on well-trained LeNet\_5 models.
The x-axis represents a dimension that is the number of neurons placed in the last feature layer of LeNet\_5 models.
As the number of neurons increases, the average angle of samples with the class direction also increases and the standard deviation will become smaller.
In our study, it is a rare event that a feature vector has a small angle (e.g., 0 degrees) with its class direction in the popular well-trained convolution architectures. 
Most of the angle between samples and the class direction is above 30 degrees, and the angle of samples with unrelated class direction is near 90.

%\BK{It will be good to have a summary statement here that tells the reader how these metrics and findings will be used later. Basically, it is an interesting observation but why should we care and how it relates to robustness.}

%
%The ortho
%In the later section, we will show that it is rare that a sample has small (e.g., 0 degree) angle between the target semantic direction and its feature representation.

%\begin{figure}[t]
%\centering
%    \includegraphics[width=\linewidth]{images/4_curse_of_dimension_1.png}
   
%    \vspace{-3mm}
 %   \caption{The average angle and standard deviation between two semantic direction of state of art model for Imagenet dataset.}
%\label{fig:model_semantic_angle}
%\vspace{-2mm}
%\end{figure}

\newpage

%\section{Acknowledgements}
%This work was performed under the auspices of the U.S. Department of Energy by Lawrence Livermore National Laboratory under Contract DE-AC52-07NA27344 (LLNL-CONF-764021).
%\input{backup.tex}

% or


%\ifCLASSOPTIONcompsoc
%\else
%\fi


\ifCLASSOPTIONcaptionsoff
  \newpage
\fi



\bibliographystyle{IEEEtran}
\bibliography{main.bib}

\begin{thebibliography}{10}
\providecommand{\url}[1]{#1}
\csname url@samestyle\endcsname
\providecommand{\newblock}{\relax}
\providecommand{\bibinfo}[2]{#2}
\providecommand{\BIBentrySTDinterwordspacing}{\spaceskip=0pt\relax}
\providecommand{\BIBentryALTinterwordstretchfactor}{4}
\providecommand{\BIBentryALTinterwordspacing}{\spaceskip=\fontdimen2\font plus
\BIBentryALTinterwordstretchfactor\fontdimen3\font minus
  \fontdimen4\font\relax}
\providecommand{\BIBforeignlanguage}[2]{{%
\expandafter\ifx\csname l@#1\endcsname\relax
\typeout{** WARNING: IEEEtran.bst: No hyphenation pattern has been}%
\typeout{** loaded for the language `#1'. Using the pattern for}%
\typeout{** the default language instead.}%
\else
\language=\csname l@#1\endcsname
\fi
#2}}
\providecommand{\BIBdecl}{\relax}
\BIBdecl

\bibitem{silver2016mastering}
D.~Silver, A.~Huang, C.~J. Maddison, A.~Guez, L.~Sifre, G.~Van Den~Driessche,
  J.~Schrittwieser, I.~Antonoglou, V.~Panneershelvam, M.~Lanctot \emph{et~al.},
  ``Mastering the game of go with deep neural networks and tree search,''
  \emph{nature}, vol. 529, no. 7587, pp. 484--489, 2016.

\bibitem{jumper2021highly}
J.~Jumper, R.~Evans, A.~Pritzel, T.~Green, M.~Figurnov, O.~Ronneberger,
  K.~Tunyasuvunakool, R.~Bates, A.~{\v{Z}}{\'\i}dek, A.~Potapenko
  \emph{et~al.}, ``Highly accurate protein structure prediction with
  alphafold,'' \emph{Nature}, vol. 596, no. 7873, pp. 583--589, 2021.

\bibitem{senior2020improved}
A.~W. Senior, R.~Evans, J.~Jumper, J.~Kirkpatrick, L.~Sifre, T.~Green, C.~Qin,
  A.~{\v{Z}}{\'\i}dek, A.~W. Nelson, A.~Bridgland \emph{et~al.}, ``Improved
  protein structure prediction using potentials from deep learning,''
  \emph{Nature}, vol. 577, no. 7792, pp. 706--710, 2020.

\bibitem{huang2019gpipe}
Y.~Huang, Y.~Cheng, A.~Bapna, O.~Firat, D.~Chen, M.~Chen, H.~Lee, J.~Ngiam,
  Q.~V. Le, Y.~Wu \emph{et~al.}, ``Gpipe: Efficient training of giant neural
  networks using pipeline parallelism,'' \emph{Advances in neural information
  processing systems}, vol.~32, 2019.

\bibitem{han2015deep}
S.~Han, H.~Mao, and W.~J. Dally, ``Deep compression: Compressing deep neural
  networks with pruning, trained quantization and huffman coding,'' \emph{arXiv
  preprint arXiv:1510.00149}, 2015.

\bibitem{liu2017learning}
Z.~Liu, J.~Li, Z.~Shen, G.~Huang, S.~Yan, and C.~Zhang, ``Learning efficient
  convolutional networks through network slimming,'' in \emph{Proceedings of
  the IEEE international conference on computer vision}, 2017, pp. 2736--2744.

\bibitem{hooker2019compressed}
S.~Hooker, A.~Courville, G.~Clark, Y.~Dauphin, and A.~Frome, ``What do
  compressed deep neural networks forget?'' \emph{arXiv preprint
  arXiv:1911.05248}, 2019.

\bibitem{liebenwein2021lost}
L.~Liebenwein, C.~Baykal, B.~Carter, D.~Gifford, and D.~Rus, ``Lost in pruning:
  The effects of pruning neural networks beyond test accuracy,''
  \emph{Proceedings of Machine Learning and Systems}, vol.~3, pp. 93--138,
  2021.

\bibitem{bulusu2020anomalous}
S.~Bulusu, B.~Kailkhura, B.~Li, P.~K. Varshney, and D.~Song, ``Anomalous
  example detection in deep learning: A survey,'' \emph{IEEE Access}, vol.~8,
  pp. 132\,330--132\,347, 2020.

\bibitem{hendrycks2019benchmarking}
D.~Hendrycks and T.~Dietterich, ``Benchmarking neural network robustness to
  common corruptions and perturbations,'' \emph{arXiv preprint
  arXiv:1903.12261}, 2019.

\bibitem{diffenderfer2021winning}
\BIBentryALTinterwordspacing
J.~Diffenderfer, B.~R. Bartoldson, S.~Chaganti, J.~Zhang, and B.~Kailkhura, ``A
  winning hand: Compressing deep networks can improve out-of-distribution
  robustness,'' \emph{CoRR}, vol. abs/2106.09129, 2021. [Online]. Available:
  \url{https://arxiv.org/abs/2106.09129}
\BIBentrySTDinterwordspacing

\bibitem{diffenderfer2021multiprize}
J.~Diffenderfer and B.~Kailkhura, ``Multi-prize lottery ticket hypothesis:
  Finding accurate binary neural networks by pruning a randomly weighted
  network,'' 2021.

\bibitem{van2008visualizing}
L.~Van~der Maaten and G.~Hinton, ``Visualizing data using t-sne.''
  \emph{Journal of machine learning research}, vol.~9, no.~11, 2008.

\bibitem{wold1987principal}
S.~Wold, K.~Esbensen, and P.~Geladi, ``Principal component analysis,''
  \emph{Chemometrics and intelligent laboratory systems}, vol.~2, no. 1-3, pp.
  37--52, 1987.

\bibitem{lecun1989optimal}
Y.~LeCun, J.~Denker, and S.~Solla, ``Optimal brain damage,'' \emph{Advances in
  neural information processing systems}, vol.~2, 1989.

\bibitem{frankle2018lottery}
J.~Frankle and M.~Carbin, ``The lottery ticket hypothesis: Finding sparse,
  trainable neural networks,'' \emph{arXiv preprint arXiv:1803.03635}, 2018.

\bibitem{ramanujan2020s}
V.~Ramanujan, M.~Wortsman, A.~Kembhavi, A.~Farhadi, and M.~Rastegari, ``What's
  hidden in a randomly weighted neural network?'' in \emph{Proceedings of the
  IEEE/CVF Conference on Computer Vision and Pattern Recognition}, 2020, pp.
  11\,893--11\,902.

\bibitem{li2020cnnpruner}
G.~Li, J.~Wang, H.-W. Shen, K.~Chen, G.~Shan, and Z.~Lu, ``Cnnpruner: Pruning
  convolutional neural networks with visual analytics,'' \emph{IEEE
  Transactions on Visualization and Computer Graphics}, vol.~27, no.~2, pp.
  1364--1373, 2020.

\bibitem{wang2019deepvid}
J.~Wang, L.~Gou, W.~Zhang, H.~Yang, and H.-W. Shen, ``Deepvid: Deep visual
  interpretation and diagnosis for image classifiers via knowledge
  distillation,'' \emph{IEEE transactions on visualization and computer
  graphics}, vol.~25, no.~6, pp. 2168--2180, 2019.

\bibitem{hohman2019s}
F.~Hohman, H.~Park, C.~Robinson, and D.~H.~P. Chau, ``S ummit: Scaling deep
  learning interpretability by visualizing activation and attribution
  summarizations,'' \emph{IEEE transactions on visualization and computer
  graphics}, vol.~26, no.~1, pp. 1096--1106, 2019.

\bibitem{8802509}
M.~Liu, S.~Liu, H.~Su, K.~Cao, and J.~Zhu, ``Analyzing the noise robustness of
  deep neural networks,'' in \emph{2018 IEEE Conference on Visual Analytics
  Science and Technology (VAST)}, 2018, pp. 60--71.

\bibitem{kahng2017cti}
M.~Kahng, P.~Y. Andrews, A.~Kalro, and D.~H. Chau, ``A cti v is: Visual
  exploration of industry-scale deep neural network models,'' \emph{IEEE
  transactions on visualization and computer graphics}, vol.~24, no.~1, pp.
  88--97, 2017.

\bibitem{sun2021certified}
J.~Sun, A.~Mehra, B.~Kailkhura, P.-Y. Chen, D.~Hendrycks, J.~Hamm, and Z.~M.
  Mao, ``Certified adversarial defenses meet out-of-distribution corruptions:
  Benchmarking robustness and simple baselines,'' \emph{arXiv preprint
  arXiv:2112.00659}, 2021.

\bibitem{mintun2021interaction}
E.~Mintun, A.~Kirillov, and S.~Xie, ``On interaction between augmentations and
  corruptions in natural corruption robustness,'' \emph{Advances in Neural
  Information Processing Systems}, vol.~34, 2021.

\bibitem{michaelis2019benchmarking}
C.~Michaelis, B.~Mitzkus, R.~Geirhos, E.~Rusak, O.~Bringmann, A.~S. Ecker,
  M.~Bethge, and W.~Brendel, ``Benchmarking robustness in object detection:
  Autonomous driving when winter is coming,'' \emph{arXiv preprint
  arXiv:1907.07484}, 2019.

\bibitem{sun2022benchmarking}
J.~Sun, Q.~Zhang, B.~Kailkhura, Z.~Yu, C.~Xiao, and Z.~M. Mao, ``Benchmarking
  robustness of 3d point cloud recognition against common corruptions,''
  \emph{arXiv preprint arXiv:2201.12296}, 2022.

\bibitem{ren2016squares}
D.~Ren, S.~Amershi, B.~Lee, J.~Suh, and J.~D. Williams, ``Squares: Supporting
  interactive performance analysis for multiclass classifiers,'' \emph{IEEE
  transactions on visualization and computer graphics}, vol.~23, no.~1, pp.
  61--70, 2016.

\bibitem{zhang2018manifold}
J.~Zhang, Y.~Wang, P.~Molino, L.~Li, and D.~S. Ebert, ``Manifold: A
  model-agnostic framework for interpretation and diagnosis of machine learning
  models,'' \emph{IEEE transactions on visualization and computer graphics},
  vol.~25, no.~1, pp. 364--373, 2018.

\bibitem{hinterreiter2020confusionflow}
A.~Hinterreiter, P.~Ruch, H.~Stitz, M.~Ennemoser, J.~Bernard, H.~Strobelt, and
  M.~Streit, ``Confusionflow: A model-agnostic visualization for temporal
  analysis of classifier confusion,'' \emph{IEEE Transactions on Visualization
  and Computer Graphics}, 2020.

\bibitem{chatzimparmpas2020stackgenvis}
A.~Chatzimparmpas, R.~M. Martins, K.~Kucher, and A.~Kerren, ``Stackgenvis:
  Alignment of data, algorithms, and models for stacking ensemble learning
  using performance metrics,'' \emph{IEEE Transactions on Visualization and
  Computer Graphics}, vol.~27, no.~2, pp. 1547--1557, 2020.

\bibitem{wang2022learning}
J.~Wang, L.~Wang, Y.~Zheng, C.-C.~M. Yeh, S.~Jain, and W.~Zhang,
  ``Learning-from-disagreement: A model comparison and visual analytics
  framework,'' \emph{IEEE Transactions on Visualization and Computer Graphics},
  2022.

\bibitem{li2020visual}
Y.~Li, T.~Fujiwara, Y.~K. Choi, K.~K. Kim, and K.-L. Ma, ``A visual analytics
  system for multi-model comparison on clinical data predictions,''
  \emph{Visual Informatics}, vol.~4, no.~2, pp. 122--131, 2020.

\bibitem{guo2017calibration}
C.~Guo, G.~Pleiss, Y.~Sun, and K.~Q. Weinberger, ``On calibration of modern
  neural networks,'' in \emph{International Conference on Machine
  Learning}.\hskip 1em plus 0.5em minus 0.4em\relax PMLR, 2017, pp. 1321--1330.

\bibitem{xia2021revisiting}
J.~Xia, Y.~Zhang, J.~Song, Y.~Chen, Y.~Wang, and S.~Liu, ``Revisiting
  dimensionality reduction techniques for visual cluster analysis: An empirical
  study,'' \emph{IEEE Transactions on Visualization and Computer Graphics},
  vol.~28, no.~1, pp. 529--539, 2021.

\bibitem{liu2014distortion}
S.~Liu, B.~Wang, P.-T. Bremer, and V.~Pascucci, ``Distortion-guided
  structure-driven interactive exploration of high-dimensional data,'' in
  \emph{Computer Graphics Forum}, vol.~33, no.~3.\hskip 1em plus 0.5em minus
  0.4em\relax Wiley Online Library, 2014, pp. 101--110.

\bibitem{blalock2020state}
D.~Blalock, J.~J. Gonzalez~Ortiz, J.~Frankle, and J.~Guttag, ``What is the
  state of neural network pruning?'' \emph{Proceedings of machine learning and
  systems}, vol.~2, pp. 129--146, 2020.

\bibitem{lee2018snip}
N.~Lee, T.~Ajanthan, and P.~H. Torr, ``Snip: Single-shot network pruning based
  on connection sensitivity,'' \emph{ICLR}, 2019.

\bibitem{molchanov2019importance}
P.~Molchanov, A.~Mallya, S.~Tyree, I.~Frosio, and J.~Kautz, ``Importance
  estimation for neural network pruning,'' in \emph{CVPR}, 2019, pp.
  11\,264--11\,272.

\bibitem{hassibi1992second}
B.~Hassibi and D.~Stork, ``Second order derivatives for network pruning:
  Optimal brain surgeon,'' \emph{Advances in neural information processing
  systems}, vol.~5, 1992.

\bibitem{sreenivasan2021finding}
K.~Sreenivasan, S.~Rajput, J.-Y. Sohn, and D.~Papailiopoulos, ``Finding
  everything within random binary networks,'' \emph{arXiv preprint
  arXiv:2110.08996}, 2021.

\bibitem{chen2020angular}
B.~Chen, W.~Liu, Z.~Yu, J.~Kautz, A.~Shrivastava, A.~Garg, and A.~Anandkumar,
  ``Angular visual hardness,'' in \emph{International Conference on Machine
  Learning}.\hskip 1em plus 0.5em minus 0.4em\relax PMLR, 2020, pp. 1637--1648.

\bibitem{liu2016large}
W.~Liu, Y.~Wen, Z.~Yu, and M.~Yang, ``Large-margin softmax loss for
  convolutional neural networks,'' \emph{arXiv preprint arXiv:1612.02295},
  2016.

\bibitem{shneiderman2003eyes}
B.~Shneiderman, ``The eyes have it: A task by data type taxonomy for
  information visualizations,'' in \emph{The craft of information
  visualization}.\hskip 1em plus 0.5em minus 0.4em\relax Elsevier, 2003, pp.
  364--371.

\bibitem{liu2018rethinking}
Z.~Liu, M.~Sun, T.~Zhou, G.~Huang, and T.~Darrell, ``Rethinking the value of
  network pruning,'' \emph{arXiv preprint arXiv:1810.05270}, 2018.

\bibitem{tulio2020beyond}
M.~Tulio~Ribeiro, T.~Wu, C.~Guestrin, and S.~Singh, ``Beyond accuracy:
  Behavioral testing of nlp models with checklist,'' \emph{arXiv e-prints}, pp.
  arXiv--2005, 2020.

\bibitem{mcinnes2018umap}
L.~McInnes, J.~Healy, and J.~Melville, ``Umap: Uniform manifold approximation
  and projection for dimension reduction,'' \emph{arXiv preprint
  arXiv:1802.03426}, 2018.

\bibitem{xenopoulos2022calibrate}
P.~Xenopoulos, J.~Rulff, L.~G. Nonato, B.~Barr, and C.~Silva, ``Calibrate:
  Interactive analysis of probabilistic model output,'' \emph{IEEE Transactions
  on Visualization and Computer Graphics}, 2022.

\bibitem{adadi2018peeking}
A.~Adadi and M.~Berrada, ``Peeking inside the black-box: a survey on
  explainable artificial intelligence (xai),'' \emph{IEEE access}, vol.~6, pp.
  52\,138--52\,160, 2018.

\bibitem{simonyan2013deep}
K.~Simonyan, A.~Vedaldi, and A.~Zisserman, ``Deep inside convolutional
  networks: Visualising image classification models and saliency maps,''
  \emph{arXiv preprint arXiv:1312.6034}, 2013.

\bibitem{adebayo2018sanity}
J.~Adebayo, J.~Gilmer, M.~Muelly, I.~Goodfellow, M.~Hardt, and B.~Kim, ``Sanity
  checks for saliency maps,'' \emph{Advances in neural information processing
  systems}, vol.~31, 2018.

\bibitem{kornblith2019similarity}
S.~Kornblith, M.~Norouzi, H.~Lee, and G.~Hinton, ``Similarity of neural network
  representations revisited,'' in \emph{International conference on machine
  learning}.\hskip 1em plus 0.5em minus 0.4em\relax PMLR, 2019, pp. 3519--3529.

\bibitem{selvaraju2017grad}
R.~R. Selvaraju, M.~Cogswell, A.~Das, R.~Vedantam, D.~Parikh, and D.~Batra,
  ``Grad-cam: Visual explanations from deep networks via gradient-based
  localization,'' in \emph{Proceedings of the IEEE international conference on
  computer vision}, 2017, pp. 618--626.

\bibitem{kim2018interpretability}
B.~Kim, M.~Wattenberg, J.~Gilmer, C.~Cai, J.~Wexler, F.~Viegas \emph{et~al.},
  ``Interpretability beyond feature attribution: Quantitative testing with
  concept activation vectors (tcav),'' in \emph{International conference on
  machine learning}.\hskip 1em plus 0.5em minus 0.4em\relax PMLR, 2018, pp.
  2668--2677.

\bibitem{verleysen2005curse}
M.~Verleysen and D.~Fran{\c{c}}ois, ``The curse of dimensionality in data
  mining and time series prediction,'' in \emph{International work-conference
  on artificial neural networks}.\hskip 1em plus 0.5em minus 0.4em\relax
  Springer, 2005, pp. 758--770.

\end{thebibliography}

\end{document}